\documentclass[aps,showpacs,nofootinbib]{revtex4}

\usepackage{amsmath,amssymb,graphicx,float,color}

\begin{document}

\title{Statistical characterization of deviations from planned flight trajectories in air traffic management}
\author{C. Bongiorno$^{1}$, G. Gurtner$^{2,3}$, F. Lillo$^{4}$, R. N. Mantegna$^{1,5,6}$, S. Miccich\`e$^{1}$}
\affiliation{$^{1}$ Universit\`a degli Studi di Palermo, Dipartimento di Fisica e Chimica, Viale delle Scienze, Ed. 18, I-90128, Palermo, Italy \\
                $^{2}$ Deep Blue srl, P.zza Buenos Aires, Roma, Italy\\
                $^{3}$ University of Westminster, 35 Marylebone Road, London NW1 5LS, United Kingdom\\
                $^{4}$ Scuola Normale Superiore, Piazza dei Cavalieri 7, Pisa, Italy\\
                $^{5}$ Central European University, Center for Network Science, Nador 9, H-1051, Budapest, Hungary \\
                $^{6}$ Central European University, Department of Economics, Nador 9, H-1051, Budapest, Hungary}

\date{\today}

\begin{abstract}


Understanding the relation between planned and realized flight trajectories and the determinants of flight deviations is of great importance in air traffic management. In this paper we perform an in-depth investigation of the statistical properties of planned and realized air traffic on the German airspace during a 28 day periods, corresponding to an AIRAC cycle.  We find that realized trajectories are on average shorter than planned ones and this effect is stronger during night-time than day-time. Flights are more frequently deviated close to the departure airport and at a relatively large angle-to-destination. Moreover, the probability of a deviation is higher in low traffic phases. All these evidences indicate that deviations are mostly used by controllers to give directs to flights when traffic conditions allow it.   Finally we introduce a new metric, termed di-fork, which is able to characterize navigation points according to the likelihood that a deviation occurs there. Di-fork allows to identify in a statistically rigorous way navigation point pairs where deviations are more (less) frequent than expected under a null hypothesis of randomness that takes into account the heterogeneity of the navigation points. Such pairs can therefore be seen as sources of flexibility (stability) of controllersÕ traffic management while conjugating safety and efficiency.

\end{abstract}

\maketitle


\section{Introduction}

In the recent literature it is possible to find many examples where network science has been applied to the air transportation system (for a review, see \cite{ZaninLillo13,CWbook}). Many studies have focused on the topological aspect of the airport network \cite{AIRP1,AIRP2,AIRP3,AIRP4,AIRP5,AIRP6,AIRP7,AIRP8,AIRP9,AIRP10,AIRP11,AIRP12}, but network science techniques can also be used to study topics more related to air traffic management \cite{ATM1,ATM2,ATM3,ATM4,ATM5,ATM6,ATM7,ATM8,ATM9,ATM10,ATM11}. In particular, one can consider different elements of the airspace like sectors and navigation points and build networks which are informative about the air traffic management \cite{plos}. In fact, differently than  the airport network, navigation point networks are more related to air traffic management problems and to safety issues.  

Here we present a study of the air traffic management procedures controlling the flow of flights occurring on top of the navigation point network. Navigation points are fixed two dimensional points in the airspace specified by latitude and longitude. The airlines must use this grid to plan each flight trajectory from departure to destination. Navigation points are also of reference for air traffic controllers who use them to solve conflicts and problems originated by unforeseen events and to rationalize and decrease the complexity of the aircraft flow. The navigation points can be viewed as a guide for airlines, but also as a burden, because flights cannot fly straight and have to find a path on this predefined grid. In fact, it is foreseen by the SESAR project \cite{conops} that navigation points will slowly disappear to allow smooth trajectories, the so-called ``business'' trajectories. However, in the present air transportation system they are crucial for air traffic controllers. In the present work we will focus on a quantitative assessment of their role in the air traffic management. 

In our study we investigate how the planned flight trajectories are modified by controllers in relationship with unforeseen events or pilots' requests. Our study is based on a metric called directional-fork, or {\em{di-fork}}, comparing planned flight trajectories with deviated flight trajectories. By using this metric we obtain a quantitative description of the deviations of planned flight trajectories called by air traffic controllers at the level of single navigation point pairs. The activity of air traffic controllers usually concerns two main aspects: on one side they are responsible for avoiding safety events and for making the aircraft trajectories conflict-free. On the other side, whenever possible, they can issue directs that $(i)$ shorten trajectories, thus allowing for lower fuel consumption, and $(ii)$ can improve the predictability of the system. In our investigations we show that directs are the main determinants for the probability of flight trajectory deviations.

We perform a statistical validation of the navigations point pairs by comparing the observed values of the di-fork metric 
with the values expected under a null hypothesis 
of deviations occurring at randomly distributed navigation point pairs.
In other words, we investigate how the different navigation points present in a given airspace are used
by air traffic controllers over the day. 
Specifically, we detect navigation point pairs where trajectories $(i)$ are most likely to be deviated with respect to the planned ones, thus providing a ``destabilization'' of the planned trajectory, or $(ii)$ are most likely not to be deviated with respect to the planned ones, thus providing a ``stabilization'' of the planned trajectory.

The paper is organized as follows: in section \ref{data} we describe the database used in our investigation. Section \ref{statinv} deals with the statistical investigation of planned flight trajectories.
Section \ref{empirical} focuses on the statistical properties of flight deviations observed from the planned flight trajectories. Section \ref{OEFork} introduces the di-fork and the statistical validation method used to detect a set of over-expressed and under-expressed navigation point pairs. Finally, in section \ref{concl} we draw our conclusions.

\section{Data} \label{data}

Our database contains information on all the flights that, even partly, cross the ECAC airspace. Data are collected by EUROCONTROL (http://www.eurocontrol.int), the European public institution that coordinates and plans air traffic control for all Europe and were obtained as part of the {\em{SESAR Joint Undertaking}} WP-E  research project ELSA ``Empirically grounded agent based model for the future ATM scenario''. \footnote{Data can be accessed by asking permission to the legitimate owner (EUROCONTROL). The owners reserve the right to grant/deny access to data.}

Data come from two different sources. First, we have access to the Demand Data Repository (DDR) \cite{DDR} database containing all the trajectories followed by any aircraft in the ECAC airspace during 15 months -- from the $8^{th}$ of April 2010 to the $27^{th}$ of June 2011. Each 28 day time period is termed AIRAC cycle. A planned or realized trajectory is made by  a sequence of navigation points crossed by the aircraft, together with altitudes and timestamps. The typical time between two navigation points lies between 1 and 10 minutes, giving a good time resolution for trajectories. In this paper we use the ``last filed flight plans'', i.e. the so-called M1 files, which are the planned trajectories -- filed from 6 months to one or two hours before the real departure. We also use the real trajectories, i.e. the so-called M3 files, because we will compare planned and realized trajectories in order to investigate the air traffic controllers role. 

In our study we are considering commercial flights. For this reason we have selected only scheduled flights -- excluding, in particular, military flights --  using land-plane aircraft, i.e. no helicopter, gyrocopter, etc. This gives, in first approximation, the full set of commercial flights. We also excluded all flights having a duration shorter than 10 minutes and a few other flights having obvious recording data errors.

The database includes all flights in the enlarged ECAC airspace {\footnote{Countries in the enlarged ECAC space are: Iceland (BI), Kosovo (BK), Belgium (EB), Germany-civil (ED), Estonia (EE), Finland (EF), UK (EG), Netherlands (EH), Ireland (EI), Denmark (EK), Luxembourg (EL), Norway (EN), Poland (EP), Sweden (ES), Germany-military (ET), Latvia (EV), Lithuania (EY), Albania (LA), Bulgaria (LB), Cyprus (LC), Croatia (LD), Spain (LE), France (LF), Greece (LG), Hungary (LH), Italy (LI), Slovenia (LJ), Czech Republic (LK), Malta (LM), Monaco (LN), Austria (LO), Portugal (LP), Bosnia-Herzegovina (LQ), Romania (LR), Switzerland (LS), Turkey (LT), Moldova (LU), Macedonia (LW), Gibraltar (LX), Serbia-Montenegro (LY), Slovakia (LZ), Armenia (UD), Georgia (UG), Ukraine (UK).}} even if they departed and/or landed in airports external to the enlarged ECAC airspace.

The other source of information are the NEVAC files. NEVAC files \cite{NEVAC} contain all the elements allowing the definition (borders, altitude, relationships, time of opening and closing) of the elements of airspaces, namely airblocks, sectors, airspaces (including Flight Information Region, National Airspace, Area Control Center, etc.). The active elements at a given time constitute the configuration of the airspace at that time. Thus, they give the configuration of the airspaces for an entire AIRAC cycle. Here we only use the information on sectors, airspaces and configurations to rebuild the European airspace. Specifically, at each time we have the full three dimensional boundaries of each individual sector and airspace in Europe. All this information have been gathered in a unique database, using MySQL, in order to allow fast crossed queries. 

Our investigations are mainly performed considering the flights relative to the AIRAC 334, i.e. the AIRAC starting on May 6, 2010 and ending on June 2, 2010. Data relative to other AIRACs are considered in order to check the stability of our results. We only consider flights that cross the German airspace, which is one of the European regions with the highest levels of air traffic. Specifically, we select from our database all airspace portions labeled with an ICAO code starting with ED. This would imply that small portions of the airspace of Belgium and Netherlands, mainly at high altitudes, are also included in our analyses. The boundaries of the considered airspace are shown in Fig. \ref{fig:localizationFN3} below. Moreover, to focus our analysis on the en-route phase of each flight, we filter the trajectories retaining only the portion at an altitude higher than 240 FL. Time of the day is always expressed in UTC. Finally, data do not include Saturdays and Sundays in order to avoid weekly seasonality effects.

\begin{figure} [H]
\centering
                \includegraphics[width=0.45\textwidth]{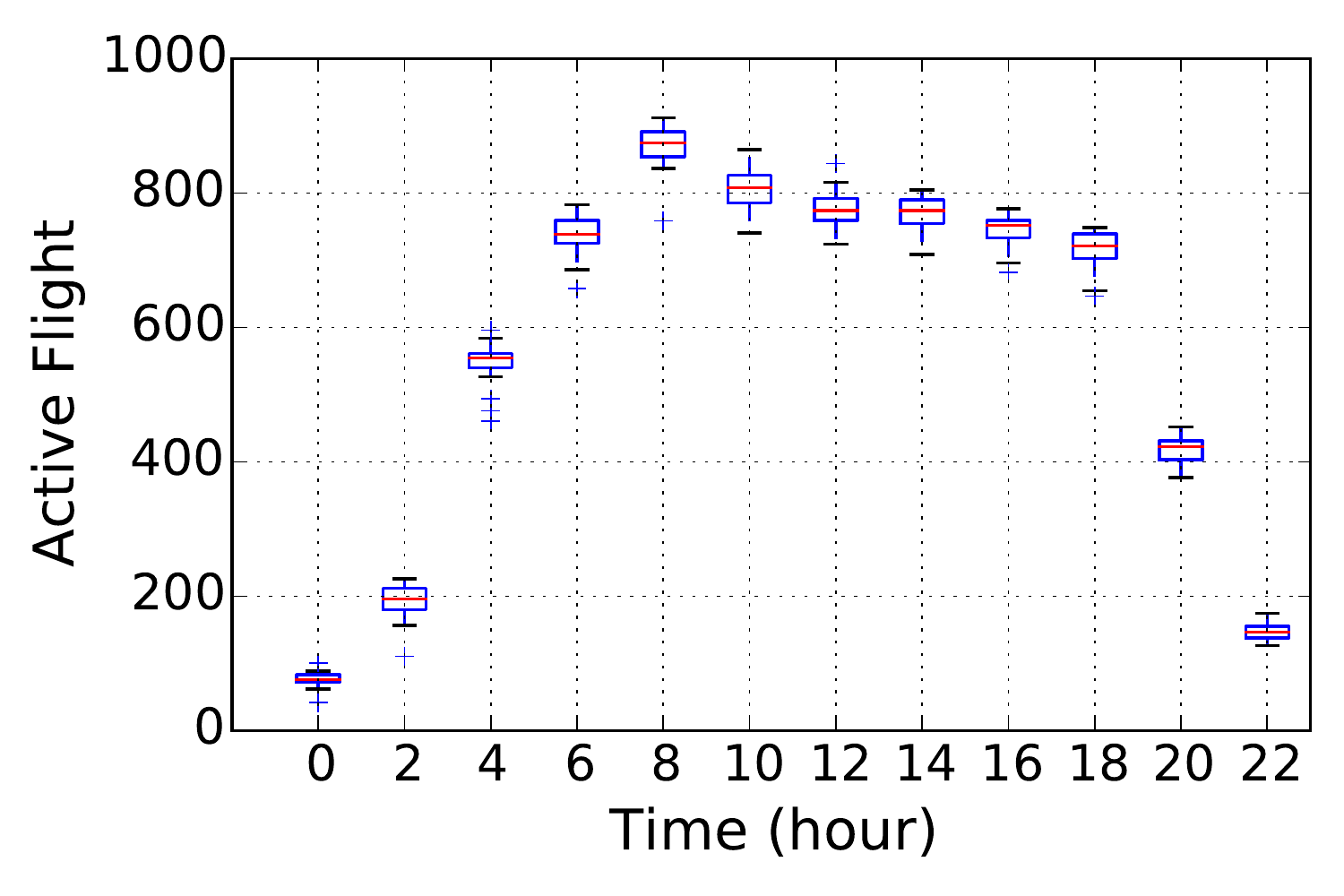}
                \includegraphics[width=0.45\textwidth]{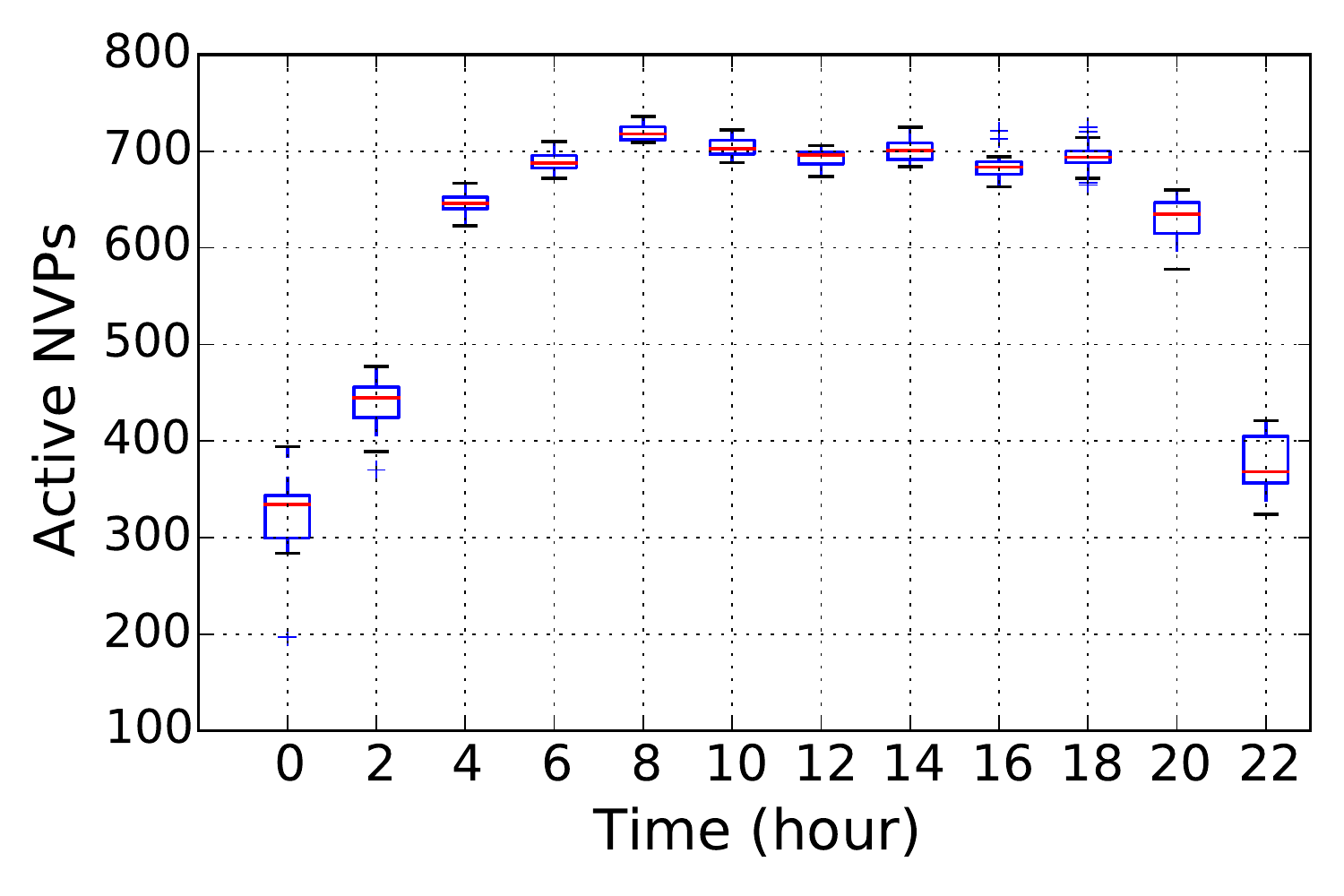}
                \caption{Box plots of the  daily number of flights (left panel) and number of active navigation points of the planned flight trajectories (right panel) for different time windows of the day. Data refer to the whole 334 AIRAC and the German (ED) airspace and are binned in two hour intervals.} \label{fig:data}
\end{figure}

In the left panel of Fig. \ref{fig:data} we show the box plot of the daily number of active flights in the different hours of the day. An intraday pattern is clearly recognizable, with many flights during day-time and almost ten times less flights during the night.  In the right panel of Fig. \ref{fig:data} we show the number of active navigation points in the planned trajectories at different hours of the day. A navigation point is active in a given time interval if at least one flight is scheduled to pass through it in that interval.  Also in this case one can see that significantly less navigation points are used during the night.

\section{Statistical properties of the length of planned and realized trajectories} \label{statinv}

When planning their flights, the airlines have to take into account many different constraints that give rise to trajectories quite different the one from the other, even for the same origin-destination airport pair in the same day. As a result, any  investigation of flight trajectories  aiming at detecting statistical regularities in the deviations occurring in the realized flight trajectories with respect to the planned ones, must take into account such heterogeneity. To this end, we present first some simple statistical facts aiming at having an overview of the data. This is the starting point of our analysis. We seek to understand under which conditions the controllers are using specific navigation points with respect for instance to the traffic conditions.

First, we associate each flight trajectory with a timestamp defined as $(i)$ the entrance time in the German airspace for flights coming from different airspaces or $(ii)$ the first time when the aircraft reaches 240 FL for flights departing inside the ED airspace. In this way, we are able to see that different types of flights are present during day and night. To show this, Table \ref{tab:length} displays the mean and the 2.5$\%$ and 97.5$\%$ percentile of the distribution of flight length (in km) during day (6:00 am - 9:00 pm) and night (9:00 pm - 6:00 am). The values shown in Table \ref{tab:length} refer to the whole flight trajectory. We consider separately planned flight trajectories  and realized flight trajectories. The first observation is that night-time flights have flight trajectories that are typically much longer than day-time flights, both for planned and realized trajectories.
\begin{table} [H]
\centering
\begin{tabular}{|c||c|c|}
\hline
                   & Planned   (km)         & Realized   (km)        \\
\hline 
\hline
Day            & $2831 \, (627,12983)$          & $2822 \, (616,12954)$                 \\
\hline
Night          & $3698 \, (655,13030)$          & $3687 \, (628,13037)$                \\
\hline
\end{tabular} 
\caption{ Average length of the planned and realized flight trajectories during day (6:00 am - 9:00 pm) and night (9:00 pm - 6:00 am). The numbers in round brackets are the values corresponding to the 2.5$\%$ and 97.5$\%$ percentile of the length distribution. All values are expressed in km. Data refer to the whole 334 AIRAC and the German (ED) airspace.}
\label{tab:length}
\end{table}

\begin{figure} [H]
\centering
                \includegraphics[width=0.45\textwidth]{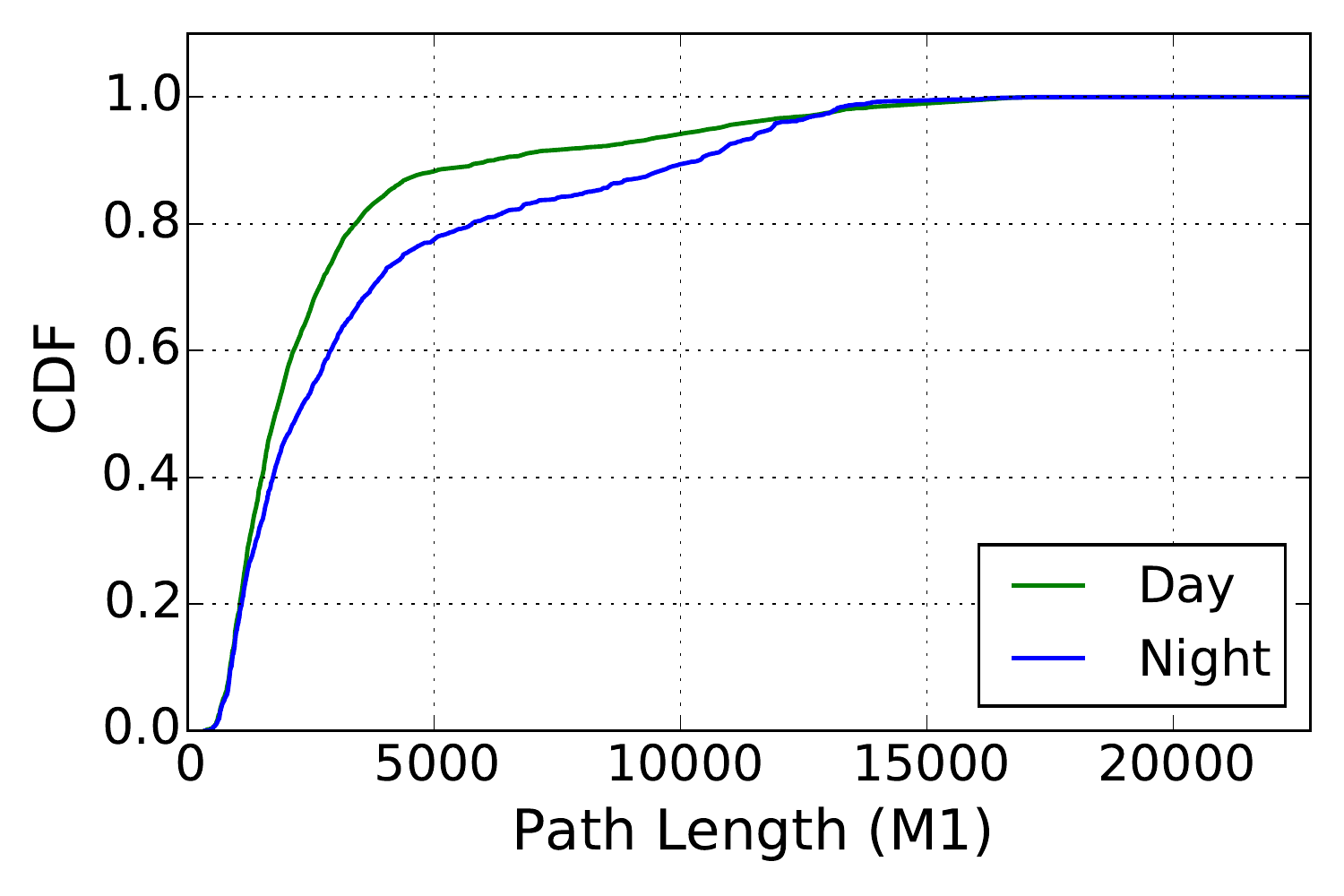}
                \caption{Cumulative distribution function of the length of planned flight trajectories in the 334 AIRAC during day (green line) and night (blue line). Data refer to the whole 334 AIRAC and the German (ED) airspace.} \label{fig:lengthdistr}
\end{figure}

This claim is confirmed by comparing the distribution function of the planned length trajectories during day and night time shown in Fig. \ref{fig:lengthdistr}. Similar results are observed when investigating the realized trajectories. 

\begin{figure} [H]
\centering
                {\includegraphics[width=0.45\textwidth]{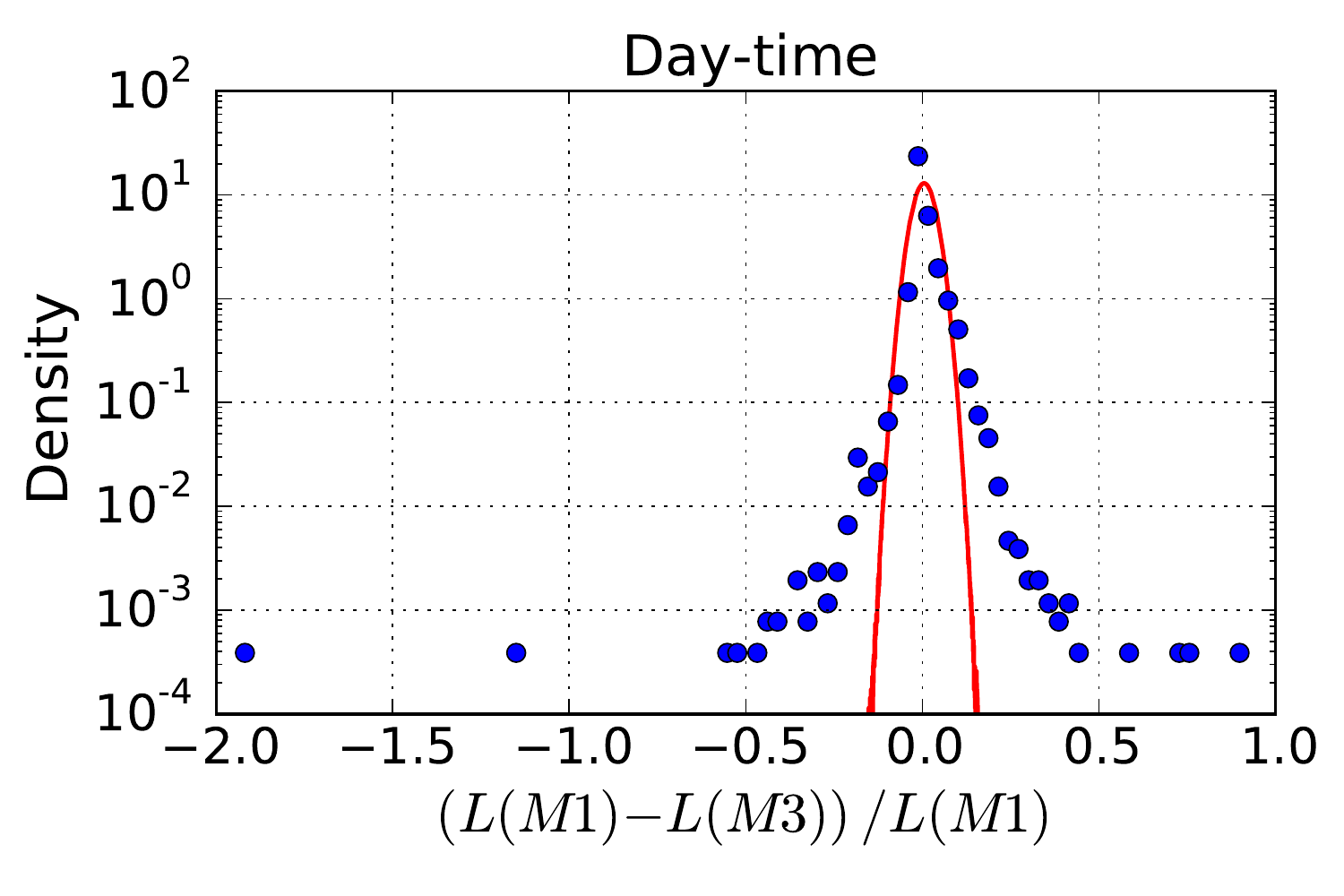}
                 \includegraphics[width=0.45\textwidth]{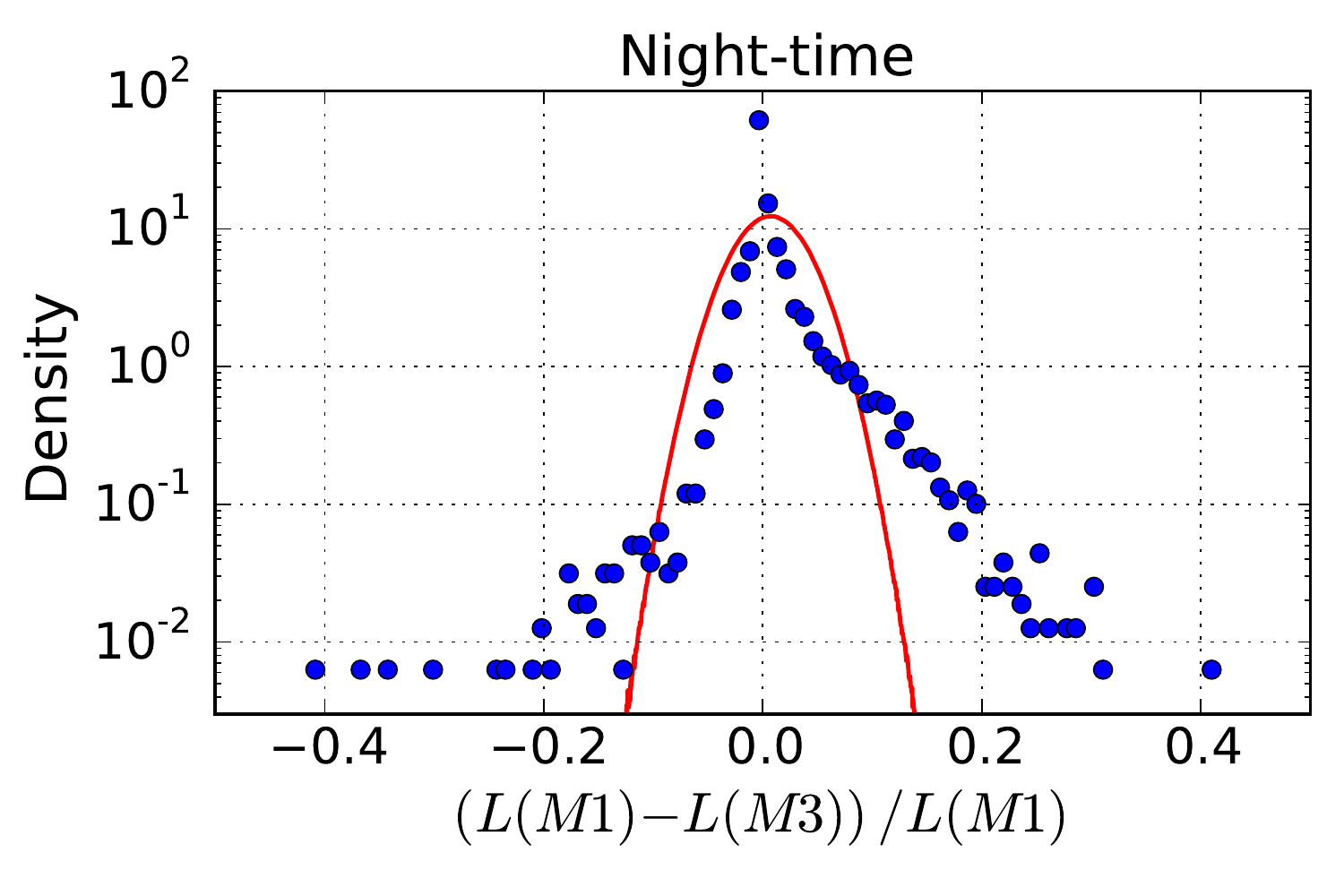}}\\
                \caption{Semilogarithmic plot of the probability density function of the relative difference between the length of the planned and realized trajectory of flights in the German airspace during the  334 AIRAC for different parts of the day: day-time from 6:00 am to 9:00 pm (left panel), and night-time from 9:00 pm to 6:00 am (right panel). The red lines are Gaussian density with mean and variance matching those of the data.} \label{fig:lengthdiffdensity}
\end{figure}

We then compare the difference in length between planned and realized trajectory, considering separately day and night. Table \ref{tab:length} shows no appreciable difference, but this is mainly due to the large heterogeneity of flight length. To have a comparison for each flight, in Figure \ref{fig:lengthdiffdensity} we show the probability density function of the fractional difference between the length of planned (labeled as {\it L(M1)}) and realized (labelled as {\it L(M3)}) flight trajectories during day (left) and night (right) time. For comparison we also show a normal distribution with the same mean and variance as the data. During day-time the distribution is almost symmetric, even if the left tail is slightly fatter than the right one, indicating a larger probability of {\it longer} realized trajectories ({\it L(M1) $<$ L(M3)}). On the contrary, for night-time flights, realized trajectories are more likely to be {\it shorter} than the planned trajectories  ({\it L(M1)$>$ L(M3)}), as indicated by the fatter right tail. In any case, empirical data show tails much fatter than Gaussian ones. 

These results indicate that the difference in planned and realized flight length has different characteristics during day and night. We now investigate whether this difference is associated with a better management of the traffic during night-time due, for example, to the presence of minor constraints in low traffic conditions. 

In order to do this, we use the ``flight efficiency'' Key Performance Indicator, as defined by the Performance Review Commission in \cite{PRR}. This efficiency is related to the best path that a flight can follow in theory, without taking into account winds. Indeed, the efficiency is obtained by comparing the planned or realized length of the $i$-th flight trajectory with the length of the shortest path between its origin airport ($O^i$) and destination airport ($D^i$). Specifically, the Efficiency for a single flight $i$ is:
\begin{equation}
                          E_i =\frac{\ell_i}{ \ell^{nvp}_i } \label{eff}
\end{equation}
where $\ell_i$ is the shortest distance between origin and destination (i.e. the grand circle) associated with the flight $i$, while $\ell^{nvp}_i$ is the length measured along its planned or realized trajectory specified by the series of successive  navigation points crossed and defining the flight trajectory. This variable takes positive values smaller than or equal to unity. In Table \ref{fig:effall} we show  the average $\mu_E$ (third and sixth columns), confidence intervals (fourth and seventh columns) and standard deviations $\sigma_E$ (fifth and eighth columns) of the efficiency computed in different time windows of the day, and considering planned and realized flight trajectories separately. Averages are taken across all investigated flights. The  confidence intervals are evaluated with a bootstrap procedure \footnote{For each two-hour time window, we store the efficiency values computed for each flight in an $E_t$ array. We therefore obtain $12$ $E_t$ arrays. To compute the confidence interval of the average efficiency for each $E_{t}$ we create $1000$ bootstrap copies $E^*_{t}$ by a sampling with replacement of the element of $E_{t}$. For each bootstrap replica of data we compute the average thus obtaining a distribution of  average values $\langle E^*_{t}\rangle$. The confidence intervals  correspond to $95\%$  respectively associated with the $2.5$ and $97.5$ percentile of the average and standard deviation distributions.}. An important result is that for all time windows the average efficiency of the planned trajectories is always smaller than the efficiency of realized trajectories. This suggests  that the air traffic controllers play an important role in increasing the system's performances.  Night-time flights (in particular during the time interval from 8:00 pm to 4:00 am) are on average more efficient than day-time flights. Moreover, the gain of average efficiency obtained in the realized trajectories is systematically larger during night-time. It is worth mentioning that the definition of Eq. \ref{eff} is a standard definition also used in Ref. \cite{PRR}. However, while we reported in the table the averages and standard deviations of the efficiency ratios, in the ATM domain it is more customary to deal with the ratio between the average values $\langle \ell_i \rangle$ and $\langle \ell^{nvp}_i \rangle$.
\begin{table}
\centering
\begin{tabular}{||c||c||c|c|c||c|c|c||}
\hline
 Time	& \# Flight	        & $\mu_E$	&	CI		& $\sigma_E$		& $\mu_E$		& CI			& 	$\sigma_E$\\
		& 			& (Planned)	& (Planned)	&	(Planned)	&	(Realized)	&	(Realized) 	& 	(Realized)	 \\
\hline 
\hline
$(0,2)$ & $1324$ &$0.9271$ & $(0.9236,0.9305)$  & $0.0640$  & $0.9382$ & $(0.9358,0.9406)$  & $0.0446$   \\ 
$(2,4)$ & $3509$ &$0.9382$ & $(0.9365,0.9398)$  & $0.0497$  & $0.9434$ & $(0.9420,0.9448)$  & $0.0425$   \\ 
$(4,6)$ & $10122$ &$0.9194$ & $(0.9180,0.9208)$  & $0.0730$  & $0.9239$ & $(0.9226,0.9253)$  & $0.0673$   \\ 
$(6,8)$ & $11650$ &$0.9131$ & $(0.9118,0.9144)$  & $0.0707$  & $0.9189$ & $(0.9171,0.9212)$  & $0.1131$   \\ 
$(8,10)$ & $14252$ &$0.9196$ & $(0.9184,0.9207)$  & $0.0687$  & $0.9234$ & $(0.9223,0.9245)$  & $0.0670$   \\ 
$(10,12)$ & $12663$ &$0.9239$ & $(0.9228,0.9250)$  & $0.0631$  & $0.9271$ & $(0.9260,0.9282)$  & $0.0651$   \\ 
$(12,14)$ & $12526$ &$0.9206$ & $(0.9193,0.9218)$  & $0.0693$  & $0.9239$ & $(0.9228,0.9250)$  & $0.0647$   \\ 
$(14,16)$ & $12485$ &$0.9208$ & $(0.9197,0.9220)$  & $0.0655$  & $0.9252$ & $(0.9242,0.9263)$  & $0.0600$   \\ 
$(16,18)$ & $11550$ &$0.9187$ & $(0.9174,0.9200)$  & $0.0707$  & $0.9220$ & $(0.9208,0.9232)$  & $0.0656$   \\ 
$(18,20)$ & $11770$ &$0.9120$ & $(0.9107,0.9133)$  & $0.0722$  & $0.9188$ & $(0.9176,0.9200)$  & $0.0662$   \\ 
$(20,22)$ & $6026$ &$0.9189$ & $(0.9170,0.9208)$  & $0.0736$  & $0.9300$ & $(0.9285,0.9315)$  & $0.0602$  \\ 
$(22,24)$ & $1857$ &$0.9226$ & $(0.9194,0.9256)$  & $0.0683$  & $0.9325$ & $(0.9298,0.9352)$  & $0.0586$   \\ 
\end{tabular} 
\caption{Average $\mu_E$ and standard deviation $\sigma_E$ of the planned and realized flight trajectories together with their 95\% confidence intervals (CI) during different time windows of the day. Confidence intervals are obtained with a bootstrap procedure. See the text for more details about the bootstrap procedure. The values reported within round brackets refer to the $2.5$ and $97.5$ percentile of the average and standard deviation distributions, respectively. Data refer to the whole 334 AIRAC and the German (ED) airspace.} \label{fig:effall}
\end{table}

The difference of the average flight trajectory efficiency between night and day and between planned and realized trajectories shows  the relevance of the navigation points structure in determining the choices of both the airlines and the air traffic controllers in the planning and management of flight trajectories. Hereafter we investigate the statistical regularities associated with the modifications of planned flight trajectories originated by the interactions between air traffic controllers and pilots. In fact,  we believe that a sound statistical characterization of the deviations occurring in the realized flight trajectories might be better performed focusing on the modifications occurring at the level of navigation points.

\section{Statistical characterization of flight trajectory deviations}   \label{empirical}

We first study whether the deviations from the planned flight trajectory occur in specific regions of the trajectories or if rather they are uniformly distributed over the trajectory. To this end, for each flight $f^i$ scheduled between the origin airport $O^i$ and the destination airport $D^i$ and for each navigation point $P_j^i$ of the planned flight trajectory, we compute the distance $d_j^i$ of the navigation point $j$ from the destination airport $D^i$. Such distance is computed along the planned flight trajectory. We then normalize such distance dividing it by the total planned length $\ell^i$ of the flight, i.e. the distance from $O^i$ to $D^i$ measured along the flight trajectory. 
The normalized distance ${\hat{d}}_j^i$ is obtained as ${\hat{d}}_j^i=d_j^i/\ell^i$. 

\begin{figure} [H]
\centering
                \includegraphics[width=0.45\textwidth]{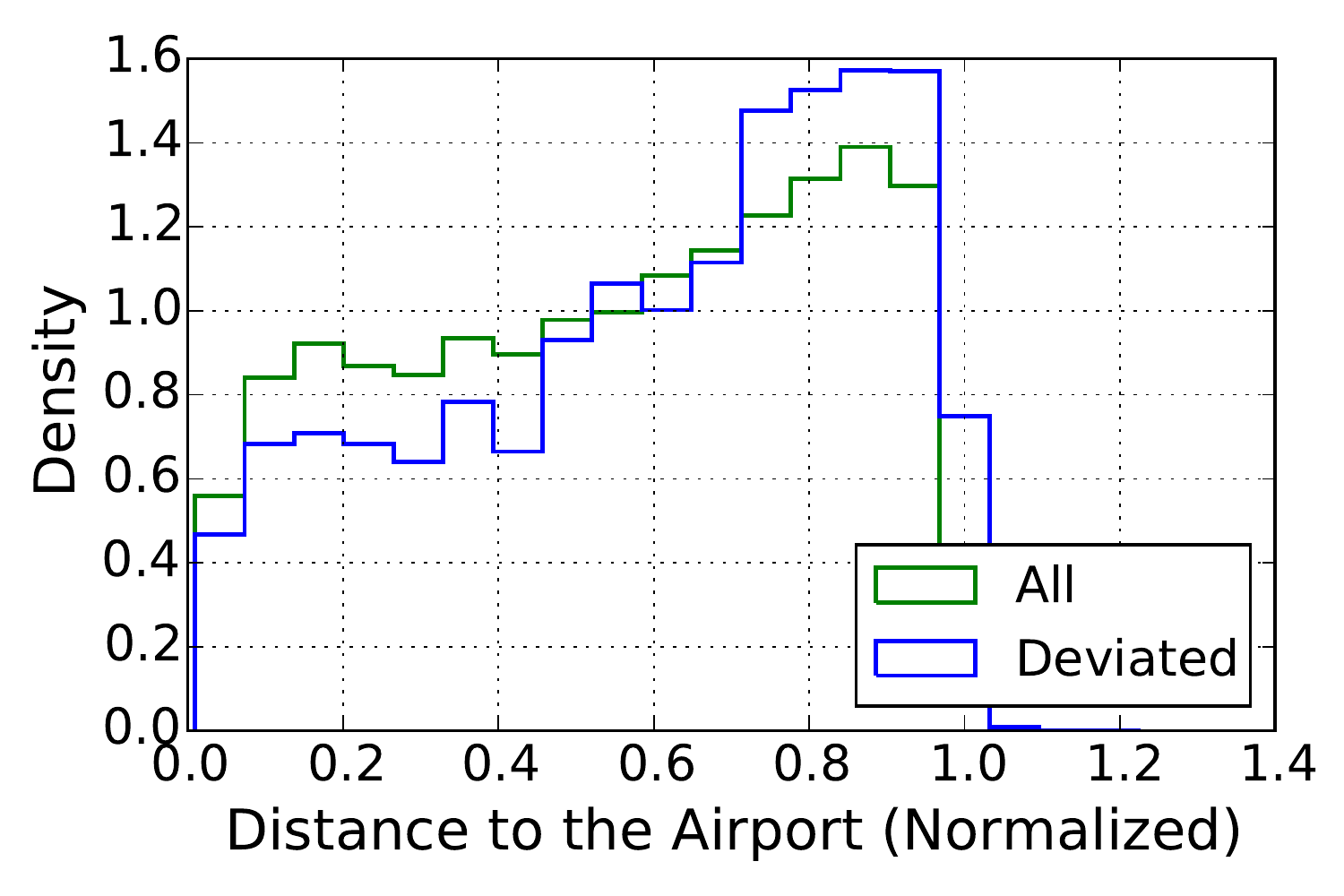}
                \caption{Probability density function of the normalized distance from the arrival airport of the navigation point where a deviation occurs (blue line). As a comparison, the green line is the probability density function of the normalized distance from the arrival airport of all navigation points crossed by the flight. Normalization is obtained by dividing the distance along the trajectory by the flight length, thus $0$ corresponds to the arrival airport and $1$ to the departing airport. Data refer to the whole 334 AIRAC and the German (ED) airspace.} \label{fig:EDdevairpdist}
\end{figure}

We call {\em{deviation}} the event in which an aircraft passing over a certain navigation point present in the planned M1 trajectory does not go to the next one as specified in the planned (M1) trajectory. In Fig. \ref{fig:EDdevairpdist} we show the probability density function of  ${\hat{d}}_j^i$ for all the navigation points and for those where we detect deviations from the planned flight trajectory\footnote{To be more precise, if an aircraft after being deviated returns back to its planned trajectory and subsequently it is deviated again, then for such flight trajectory we count two deviations.}. One can see by direct inspection that the two distributions are quite different from each other thus suggesting the idea that the navigation points where deviations occur are not randomly distributed along the trajectory. In particular, the figure shows that there are more deviations far from the destination airport. The figure thus shows that deviations frequently occurs close to the beginning of the flight trajectory, thus indicating that they might not occur to recover from accumulated en-route delay. Note that if the deviations from the planned trajectories were only the consequences of traffic regulations, the distributions would most likely be very close to each other, except for some special configuration of the airspace.

\begin{figure} [H]
\centering
                \includegraphics[width=0.18\textwidth]{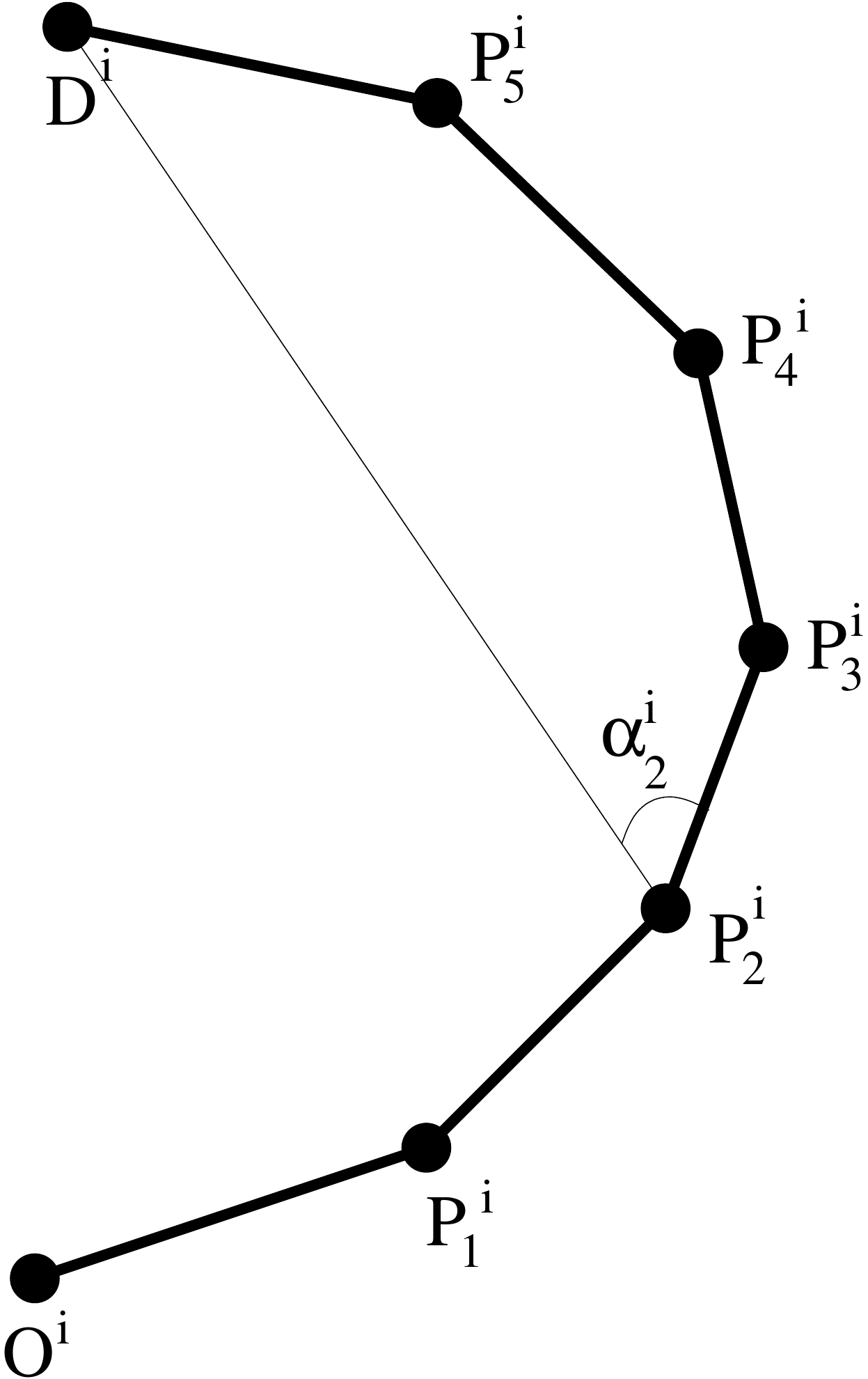} \hspace{0.8truecm}
                \includegraphics[width=0.45\textwidth]{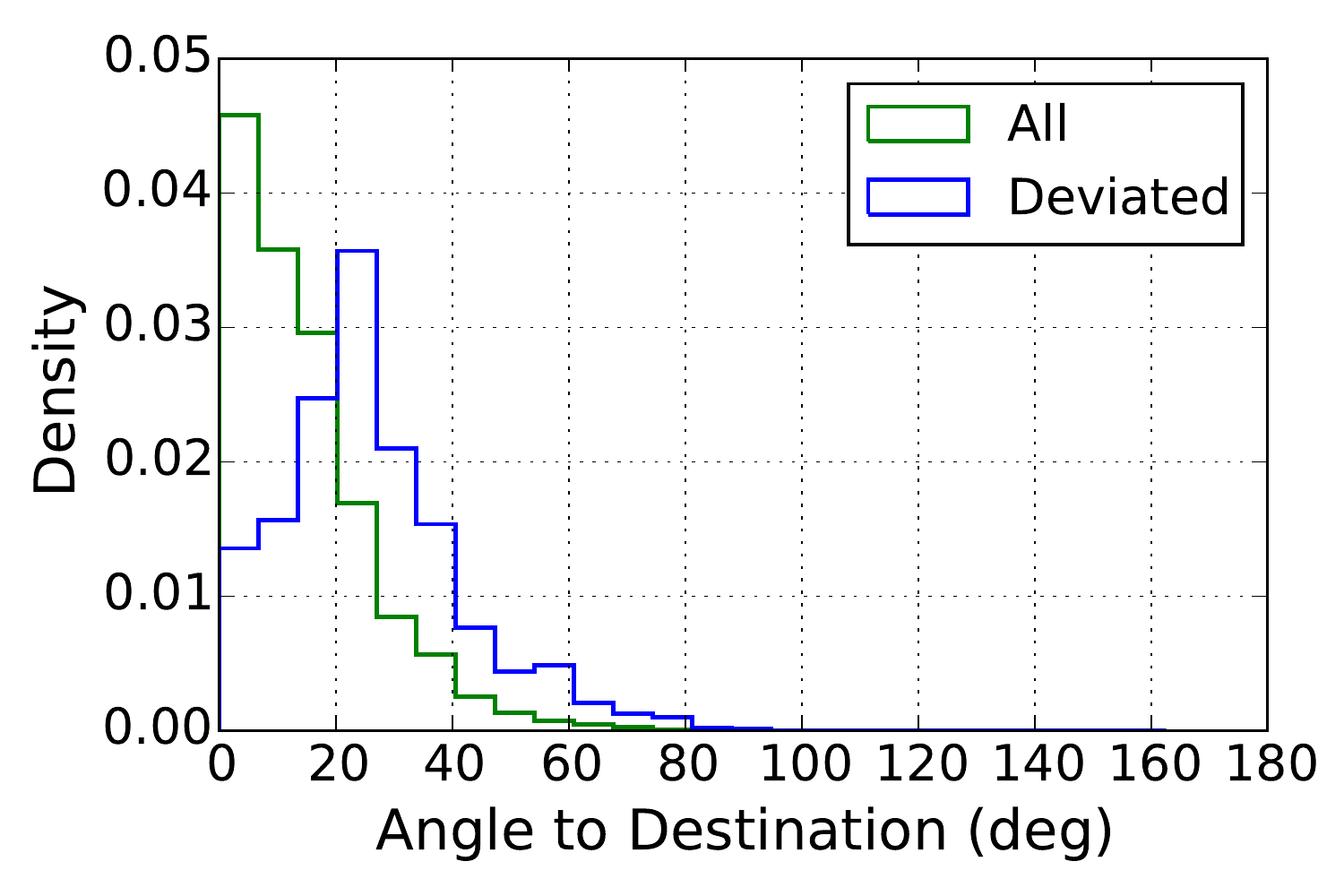}
\caption{The left panel illustrates our definition of angle-to-destination. The right panel shows the probability density function of the angle-to-destination variable estimated at a navigation point where a deviation occurs (blue line) and for all navigation points (green line). Data refer to the whole 334 AIRAC and the German (ED) airspace.} \label{fig:EDdevairpangle}
\end{figure}

We perform a similar analysis on an angle-to-destination estimator. Specifically, for each flight $f^i$ planned between the origin airport $O^i$ and the destination airport $D^i$ the angle-to-destination of  a navigation point $P_j^i$ is the angle $\alpha_j^i$ between the segment connecting the two consecutive navigation points $P_j^i$ and $P_{j+1}^{i}$ and the segment connecting $P_j^i$ with $D^i$, see the left panel of Fig. \ref{fig:EDdevairpangle}.  In the right panel of Fig. \ref{fig:EDdevairpangle} we show the probability density function of $\alpha_j^i$ for (i) all the navigation points and (ii) conditioned on  navigation points where we detect deviations from the planned trajectory. Also in this case the two distributions are quite different from each other, thus supporting the conclusion that the navigation points where deviations occur are not randomly distributed along the flight trajectory. It is also worth mentioning that, differently from the case of flight length, there exists a typical angle of $20^\circ \div 25^\circ$ for which deviations occur preferentially. One can view this angle as the typical angle after which a deviation is needed, since the aircraft is following a direction not in line with the destination airport. Note also that small angles are not represented, probably because flights in line with the destination airport do not need to be deviated. Hence, this fact advocates a major role of the directs in the cause of deviations, rather than traffic regulations.

\begin{figure} [H]
\centering 
                 \includegraphics[width=0.45\textwidth]{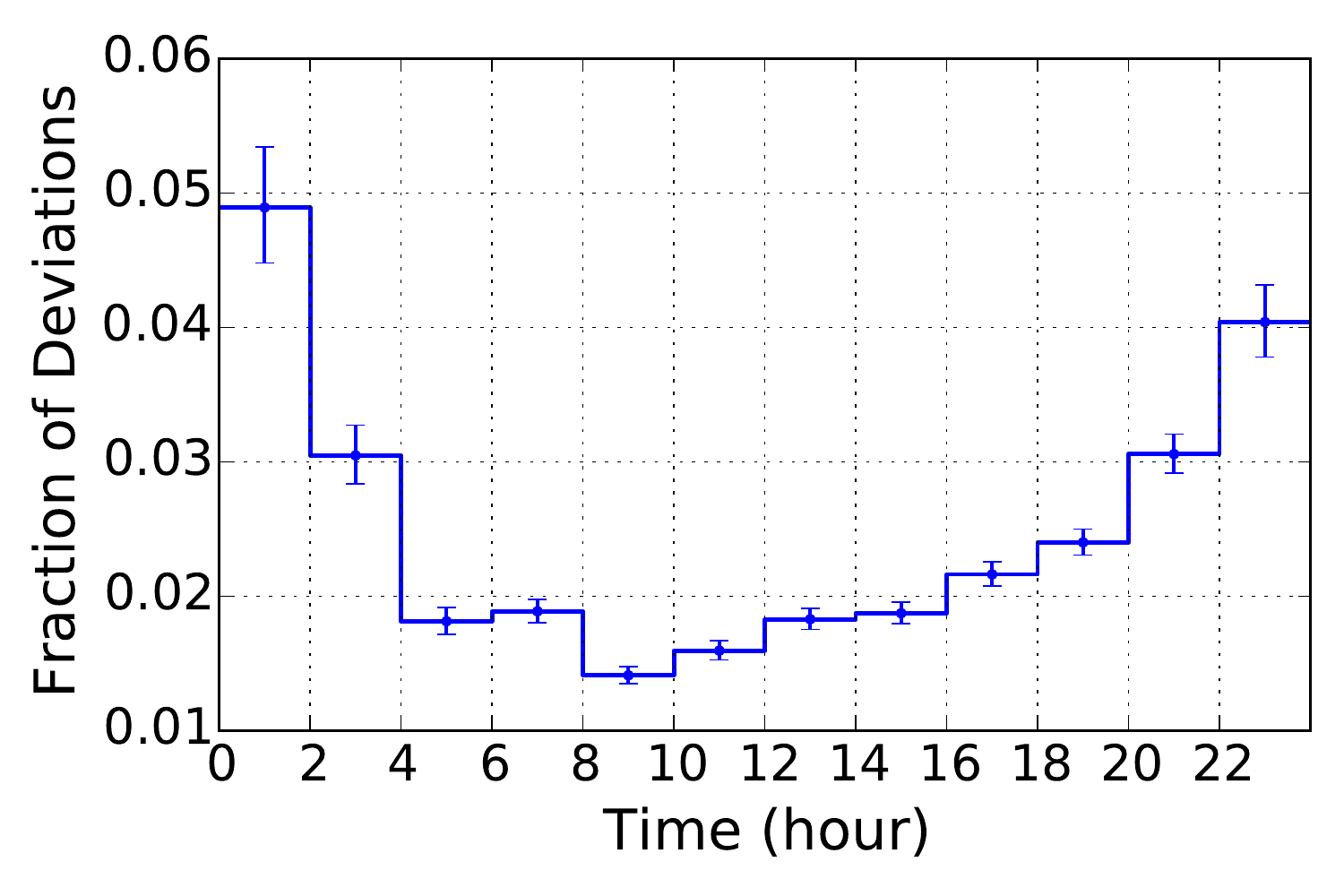}
                 \includegraphics[width=0.45\textwidth]{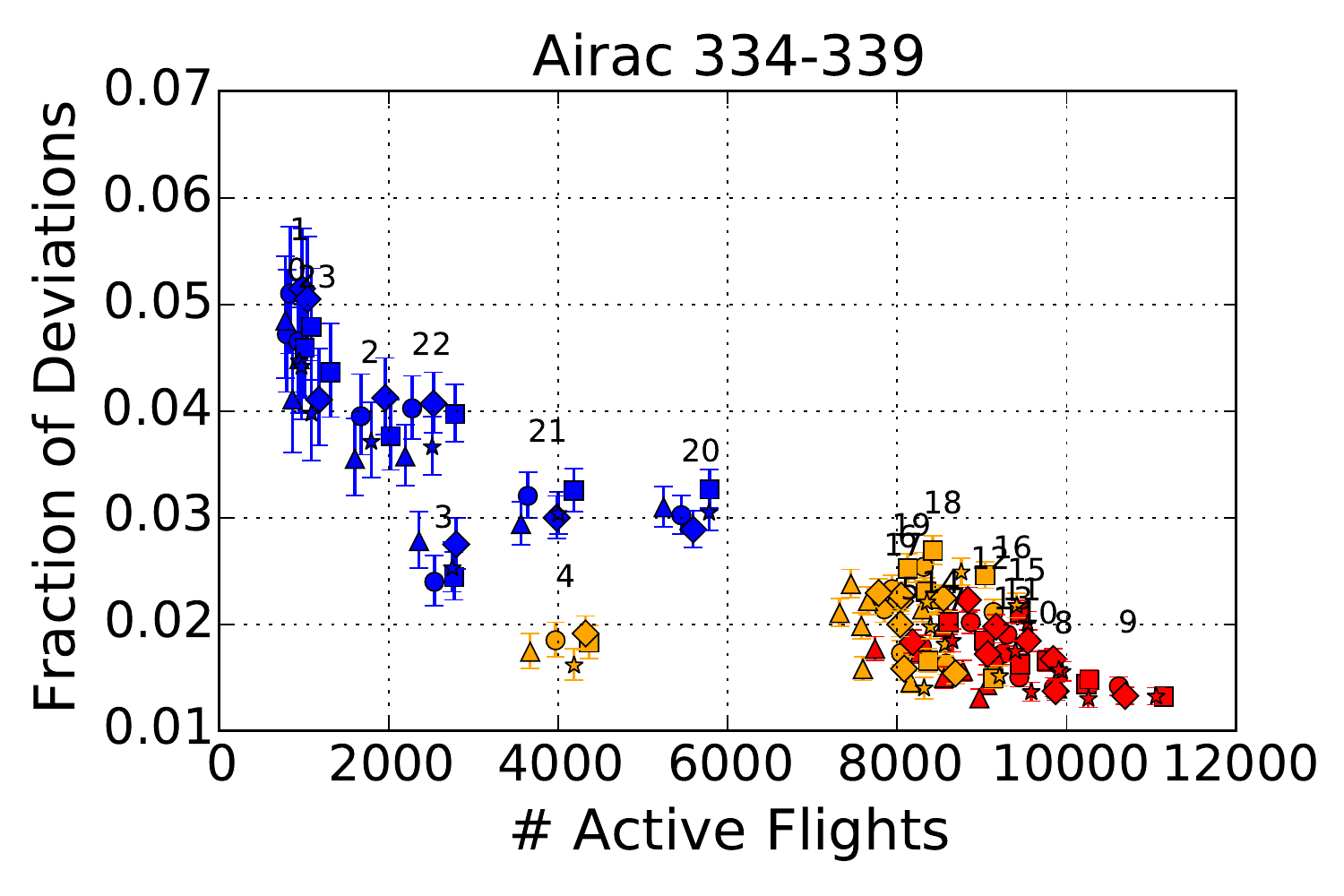}
                 \caption{In the left panel it is shown the ratio between the  number of deviations from the planned trajectory and number of possible deviations  as a function of the hour of the day. The vertical error bars represent the Wilson score interval. See the text for more details. The right panel shows such fraction of deviations  from the planned trajectory as a function of the number of active flights in different hours of the say. The number indicate the hour of the day. Different symbols refer to different AIRACs, from AIRAC 334 to AIRAC 338. Data refer to the German (ED) airspace.} \label{fig:EDdev}
\end{figure}

A third aspect we want to emphasize is that trajectory deviations do not occur in a uniform way throughout the day. Rather, we observe an intraday pattern, as shown in the left panel of Fig. \ref{fig:EDdev}. The figure shows the ratio between the observed deviations and the possible deviations in the airspace\footnote{Error bars are given by the Wilson score interval \cite{Wilson} used to associate a confidence interval to a proportion in a statistical population. The Wilson interval is an improvement over the normal approximation interval. In fact, it is more accurate even for a small number of trials and for extreme probabilities. It can be derived from Pearson's chi-squared test with two categories and it is defined as:
\begin{eqnarray}
                           \frac{1}{1 + \frac{1}{n} z^2}  \left[    \hat p + \frac{1}{2n} z^2 \pm    z \sqrt{   \frac{1}{n}\hat p \left(1 - \hat p\right) +      \frac{1}{4n^2}z^2    }  \right]
\end{eqnarray}
where $\hat{p}$ is the sample proportion observed, $n$ is the number of the trials, and $z$ is the level of significance.}. Consistently with the definition of deviations given above, the possible deviations are defined as the number of navigation points that in each M1 trajectories $(i)$ are actually crossed by the aircraft and are therefore also present in the M3 trajectories and $(ii)$ irrespective of the fact that the next planned navigation point is crossed or not. As a result the number of possible deviations coincide with the number of navigation points that are present both in the M1 and M3 flight plans. This is therefore a global metric that is not attached to a single navigation point, rather it is relative to the considered airspace. We show such ratio as a function of the time of the day in the left panel and as a function of the number of active flights in the right panel. Interestingly, the ratio of deviations is higher during night-time than during day-time. 

This observation, as well as the observation on the angles, suggests that, since traffic is lower during night, the main motivation for deviations is not the need of dealing with safety issues, but rather the possibility of issuing directs that will shorten the flight trajectories. This is confirmed by the results summarized in the right panel of Fig. \ref{fig:EDdev} where we plot the fraction of deviations as a function of the total number of aircraft present in the considered time window, averaged over an entire AIRAC. Each point refers to an hourly time-window and different symbols refer to different AIRACs, from 334 to  338.

An inverse relation between the fraction of deviated flights and the number of active flights is observed. During night-time (blue points) the traffic is lower and the fraction of deviations is larger, while the opposite is true during day-time. The figure therefore indicates that there exists an anti-correlation between fraction of deviations and traffic. When using the Pearson correlation coefficient we estimate a statistically significant correlation of -0.88, only slightly different from the value -0.83 obtained when we consider the Spearman correlation coefficient. This fact confirms that most of the deviations are actually directs, because safety issues are very few during the night.

\begin{figure} [H]
\centering
                  \includegraphics[width=0.45\textwidth]{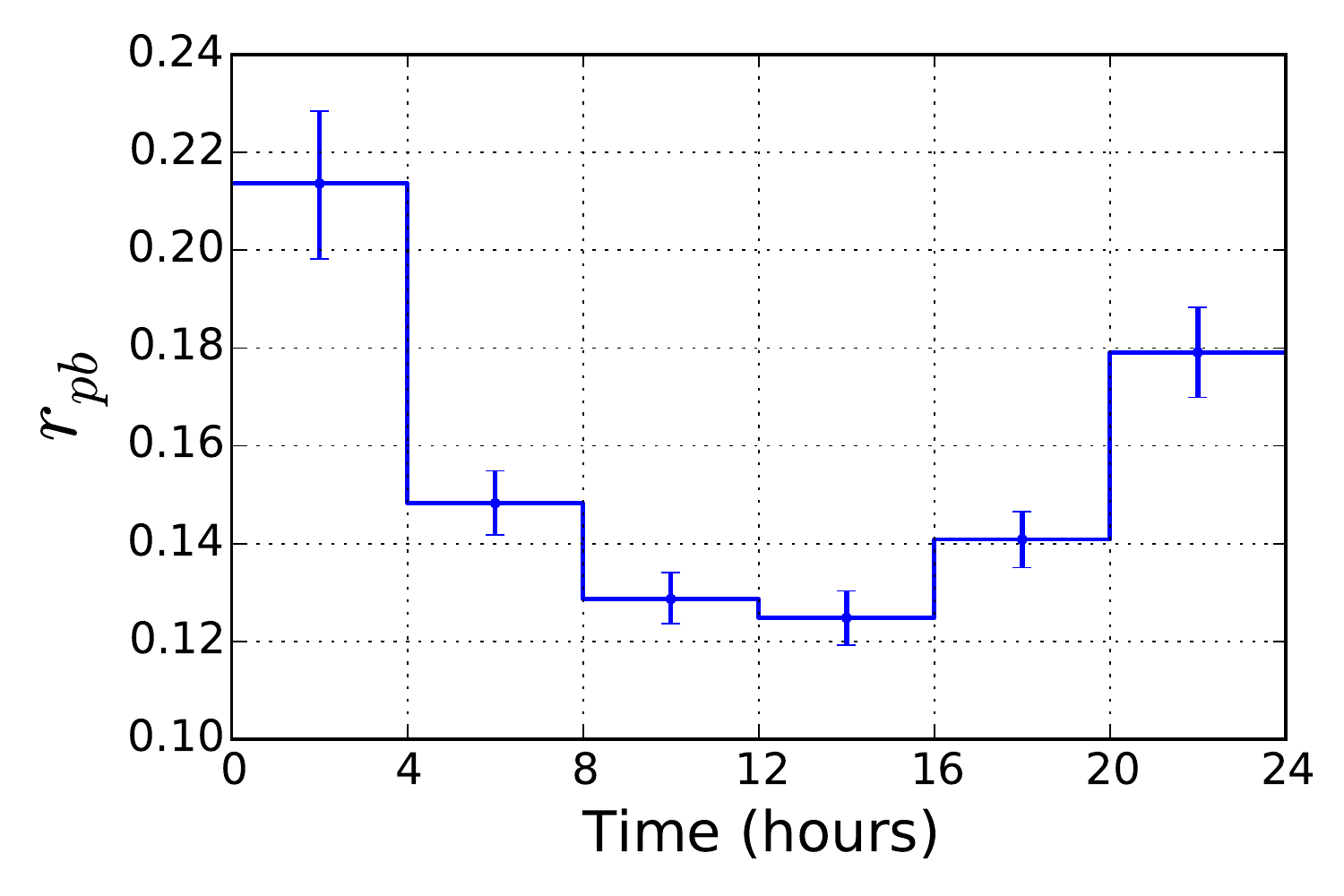}
                  \caption{Point-biserial correlation between the angle-to-destination and the categorical variable indicating the presence of a deviation of the planned flight trajectory as a function of the hour of the day. Data refer to the whole 334 AIRAC and the German (ED) airspace.} \label{fig:Rpb}
\end{figure}

Finally, Fig. \ref{fig:Rpb} shows the point-biserial correlation between the angle-to-destination for a navigation point, that is a quantitative variable, and the presence or absence of a flight deviation at the same navigation point, that is a categorical one \footnote{The point-biserial is nothing but the Pearson correlation coefficient between a vector with real entries and a boolean vector made of [0, 1]. As in Fig. \ref{fig:effall} the error bars are computed by using a bootstrap replica of data and considering a 95-percentile confidence interval.}. The figure shows that the correlation is statistically different from zero. The aim of this analysis is to assess the controllers behavior in optimizing the trajectories. In fact, the stronger the correlation, the higher the tendency of the controller in deviating aircraft. The observed intraday dynamics indicates an higher point-biserial correlation during night (from 8:00 pm to 4:00 am) rather than during the day. Again this result can be explained in terms of air traffic controllers-pilots interaction to optimize flight paths rather than interventions due to solve safety issues.  

The conclusion of the previous statistical facts is the importance of pro-active deviations rather than reactive ones. In other words, controllers usually modify the horizontal trajectories in order to speed the flights up. This leads to deviations starting early in the trajectory, triggered by high angles-to-destination and low traffic condition, usually during night. We are now interested in studying the temporal heterogeneities of the deviations, focusing on single navigation point pairs, i.e. trajectory segments.

\section{Over-expression and under-expression of flight deviations at navigation points ordered pairs} \label{OEFork}

In this section we first discuss a metric, called fork, that is used for characterizing the differences between planned and realized flight trajectories at the level of each single navigation point. The fork metric was introduced in Ref. \cite{ATMgerald}.  Let us first provide a qualitative description of it. For each flight, we consider the last navigation point which is common to the planned and the realized flight trajectory. At this point, we consider that a ``fork'' happens when the flight trajectory is deviated from the planned one.  By counting the number of flights which are deviated from the considered navigation point and by dividing it by the total number of flights flying through the navigation point in the selected time interval, we obtain a quantitative indicator of how much the navigation point is a ``source'' of deviations for the planned flight trajectories. This quantity varies between 0 and 1 and is computed for each navigation point. 

Hereafter we are providing a more formal definition. Let us consider a certain time window $\Delta t$. Let us consider a generic navigation point $P$ appearing in at least one of the realized flight trajectories. Let us call $pF_{\Delta t}(P)$ the number of flights passing through $P$ as observed in the planned flight trajectories. Let us call $dF_{\Delta t}(P)$ the number of flights passing through $P$,  as observed in the realized flight trajectories, and missing the next navigation point as indicated in the corresponding planned flight trajectory. The fork $F_{\Delta t}(P)$ is defined as the ratio $F_{\Delta t}(P)=dF_{\Delta t}(P)/pF_{\Delta t}(P)$. By construction, this metric aggregates the information on the different flight trajectories that in a certain time window $\Delta t$ are passing through $P$.

This metric already produced some interesting results presented in \cite{ATMgerald}. However, its weakness relies in the heterogeneity of trajectories which can cross a single navigation point. Indeed, controllers are managing flows, i.e. ensemble of trajectories, and for them the navigation point is a support to the flow. As a consequence, different flows crossing at a given navigation point can be managed differently. We therefore introduce a slightly different metric, where we take into account the direction of the flow as well as the navigation point itself.

\subsection{The directional fork} \label{ncm}

Let us consider pairs $C(P_j, P_k)=(P_j, P_k)$ of navigation points that are consecutively crossed according to a certain flight plan. The navigation point pairs we consider are ordered and therefore  $(P_{j}, P_k)$ and $(P_k, P_{j})$ are different pairs describing flights passing through the same pair of navigation points but moving in the opposite direction. 

Similarly to the previously mentioned fork metric, the directional fork (or {\em{di-fork}}) $\Phi_{\Delta t}(C)$ associated with an ordered navigation point pair $C$ is defined as the ratio $\Phi_{\Delta t}(C)=DF_{\Delta t}(C)/PF_{\Delta t}(C)$ where $PF_{\Delta t}(C)$ is the number of flights planned to flow through $P_j$ and $P_{k}$ in the direction from $j$ to $k$ and $DF_{\Delta t}(C)$ is the number of flights actually crossing $P_j$ and then deviated to a navigation point different from $P_{k}$ in the considered time window $\Delta t$. In other words, the first navigation point is the one crossed by the aircraft and the second one is the navigation point present in the planned flight trajectory but not present in the realized flight trajectory. This definition allows us to investigate the deviations as a function of the  different directions, and to have a more flow-based metric. It is worth emphasizing again that the di-fork metric refers to navigation point pairs, while the fork metric of Ref. \cite{ATMgerald} refers to single navigation points.

Below we investigate the capabilities of the di-fork metric in providing a statistical characterization of the deviations occurring in the flight trajectories. More specifically, we are interested in seeing how the statistical facts we found in section \ref{empirical} are present at the microscopic level, i.e. at the navigation point pair level.

\subsection{Navigation point pairs with over-expressed and under-expressed values of the di-fork metric} \label{Toeue}

Here we investigate whether the flight trajectory deviations are randomly distributed over the day or rather if they are over-expressed or under-expressed for specific navigation point pairs. This type of investigation cannot be done only in terms of the occurrence of the flight trajectory deviations because the number of flights passing through a specific navigation point pair in a given time interval is a quite heterogeneous variable. We therefore estimate the over-expression and under-expression of flight trajectory deviations by considering navigation point pairs and trying to compare the occurrences of flight trajectory deviations observed in this pair with an appropriate  null model.


In this section we investigate the navigation point pairs $C(P_j, P_k)$ for which the air traffic flow is from $P_j$ to $P_k$.  Suppose that during a specific day we have $PF_{day}$ flights with planned flight trajectories connecting $P_j$ to $P_k$ in a step. Suppose also that $DF_{day}$ is the number of flights passing through the first navigation point $P_j$ and deviating from the successive navigation point $P_k$ in the same day. Let us now define  $PF_{\Delta t}$ the flights that are planned to fly through $P_j$ and $P_k$ during an intraday time interval $\Delta t$. We can estimate what is the probability of observing a number $DF_{\Delta t}$ of flights flying through $P_j$ and then deviating from $P_k$ during the same time interval. By assuming that for each navigation point pair, the flight trajectory deviation events are independent the one from the other, a good approximation of the probability of detecting  $DF_{\Delta t}$ is given by the hypergeometric distribution\footnote{It is worth mentioning that using the hypergeometric distribution is equivalent to performing an one tail FisherÕs exact test \cite{FI1} starting from a $2 \times 2$ contingency table whose entries are $DF_{\Delta t}$ and $PF_{\Delta t}-DF_{\Delta t}$ in the first column and $DF_{day}-DF_{\Delta t}$ and $(PF_{day}-DF_{day})-(PF_{\Delta t}-DF_{\Delta t})$ in the second column \cite{FIHYP}.}:
\begin{eqnarray}
                            H( DF_{\Delta t}| PF_{day}, DF_{day},PF_{\Delta t}) = 
                                                                            \frac{\binom{DF_{day}}{DF_{\Delta t}} \, \binom{PF_{day} - DF_{day}}{PF_{\Delta t} - DF_{\Delta t}}}
                                                                                   {\binom{PF_{day}}{PF_{\Delta t}}}.    \label{E1}
\end{eqnarray}
By using this value of the probability of observing $DF_{\Delta t}$ deviated flight trajectories we can obtain for each navigation point pair $C(P_j, P_k)$ a p-value for the over-expression or the under-expression of $DF_{\Delta t}$. The probability of Eq. (\ref{E1})  allows us to associate a \emph{p}-value $p(DF_{\Delta t})$ with the actual number $DF_{\Delta t}$ of detected deviation. Specifically, for over-expression (OE) we have 
\begin{equation}
                           p_{OE}(DF_{\Delta t})=1-\sum_{X=0}^{DF_{\Delta t}-1}H( DF_{\Delta t}| PF_{day}, DF_{day},PF_{\Delta t}), \label{pvover}
\end{equation}
whereas for under-expression (UE) we have 
\begin{equation}
                        p_{UE}(DF_{\Delta t})=\sum_{X=0}^{DF_{\Delta t}}H( DF_{\Delta t}| PF_{day}, DF_{day},PF_{\Delta t}). \label{pvunder}
\end{equation}
Since we are performing this test for all possible navigation point pairs $C(P_j, P_k)$, we have to use a correction for multiple hypothesis test comparison. Specifically we use  the FDR multiple hypothesis test correction \cite{benj} therefore, after sorting the p-values in increasing order, in order to reveal over-expression or under-expression we select those navigation point pairs with p-values which are below the straight line with null intercept and slope equal to $0.01/2 N_{pair} N_t$ where $N_t=12$ is the number of used time bins and $N_{pair}$ is the number of possible pairs we tested. As an example, when testing the 334 AIRAC only we would use $N_{pair}=29076$. In the analysis presented below, where we test for over- and under-expressions over five consecutive AIRACs we will consider $N_{pair}=146112$. The factor 2 is taken into account because we want to consider both over- and under-expressions. In the results presented hereafter, we aggregate the number of temporally over-expressed and under-expressed navigation point pairs relative to the different days of an AIRAC. 

\begin{figure}[H]
\centering
                 \includegraphics[width=0.85\textwidth]{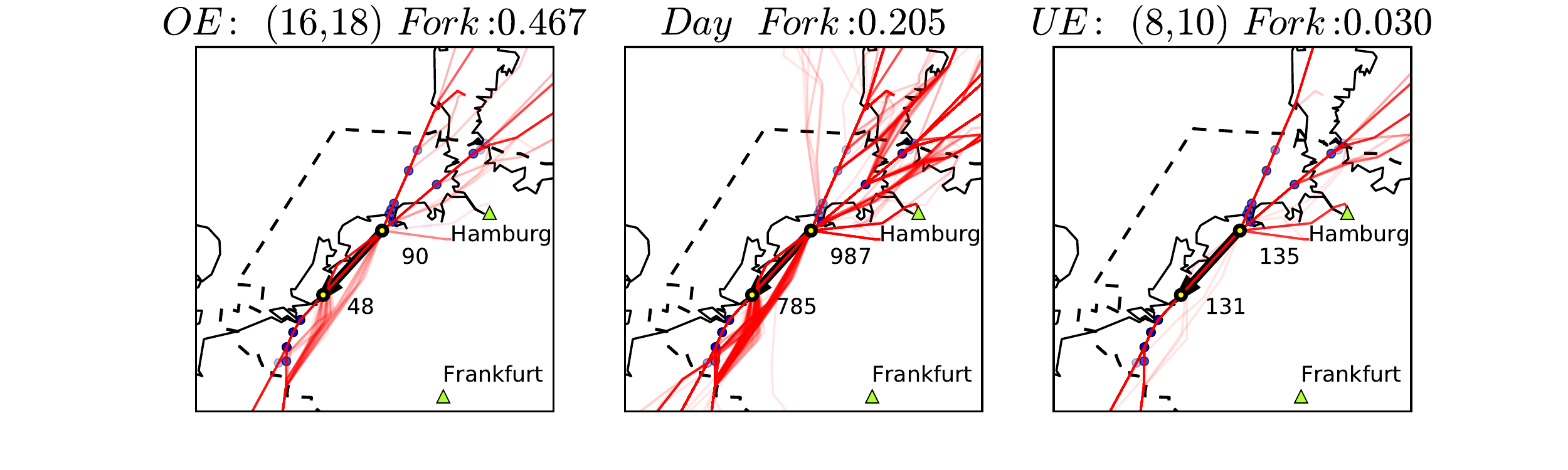}
                 \caption{The left panel shows a navigation point pair (tick red segment) where the first navigation point is crossed by 90 aircraft. However only 48 of them reach the other navigation point of the pair. This results in a di-fork value of $(90-48)/90=0.467$. The right panel shows a navigation point pair (tick red segment) where the first navigation point is crossed by 135 aircraft. and 131 of them reach the other navigation point of the pair. This results in a di-fork value of $(135-131)/135=0.030$. The central panel shows the daily behavior of the considered pair. The first navigation point is overall crossed by $987$ aircraft, with $785$ of them reaching the other navigation point of the pair. Therefore the average daily behavior is characterized by a di-fork value of $(987-785)/987=0.205$.} \label{fig:example}
\end{figure}

In Fig. \ref{fig:example} we illustrate an example of the two possible outcomes of the statistical validation procedure associated to the hypergeometric distribution described above. The left panel indicates a situation where we have an over-expressed navigation point pair (tick red segment). The first navigation point is crossed by 90 aircraft. However only 48 of them reach the other navigation point of the pair. This results in a di-fork value of $(90-48)/90=0.467$ which leads to an over-expression as a result of the comparison with the average daily behavior characterized by a di-fork value of $(987-785)/987=0.205$, see the central panel. The right panel shows the same pair in a different time window. Now there are many more aircraft crossing the two navigation points. However, the pair results to be under-expressed, given that only $135-131=4$ out of $135$ aircraft do not cross the second navigation point of the pair. The example shows that having an over-expressed or under-expressed pair does not necessarily indicate that the pair is more frequently used. Rather, the statistical validation procedure associated to the hypergeometric distribution selects the pairs where the occurrence of deviations is statistically different from the daily average.

In Table \ref{fig:EDtempOEpair} we show the number of over-expressed (second column) and under-expressed (third column) navigation point pairs in the 334 AIRAC. The fourth and fifth column show the number of navigation point pairs with at least one and five planned flights, respectively. The number of over-expressed navigation point pairs is larger in the night than during day-time, while the opposite is true for under-expressed pairs. Time windows of early morning (e.g. the 6:00 am 8:00 am time window) and of early afternoon (e.g. the 2:00 pm 4:00 pm time window) present a roughly balanced number of over-expressed and under-expressed navigation point pairs. The fifth and sixth columns indicate the number of OE and UE observed in at least one of the 5 AIRACs from 334 to 338. 
\begin{table} [H]
\centering
\begin{tabular}{||c||c|c||c|c||c|c||}
\hline
                          & number of OEs & number of UEs & number of pairs with & number of pairs with & number of OEs in & number of UEs  in \\
                          &    334 AIRAC     &   334 AIRAC     &        at least 1 flight   & at least 5 flights         & at least 1 AIRAC   & at least 1 AIRAC \\
\hline 
\hline
$[0,2]$ & $10$ & $1$ &  $876$ &  $510$  &  $19$ &  $1$ \\ \hline
$[2,4]$ & $12$ & $0$ &  $977$ &  $667$  &  $34$ &  $0$ \\ \hline
$[4,6]$ & $1$ & $0$ &  $1316$ &  $1003$  &  $6$ &  $2$ \\ \hline
$[6,8]$ & $3$ & $2$ &  $1386$ &  $1108$  &  $6$ &  $6$ \\ \hline
$[8,10]$ & $1$ & $6$ &  $1442$ &  $1157$  &  $5$ &  $17$ \\ \hline
$[10,12]$ & $0$ & $2$ &  $1426$ &  $1160$  &  $1$ &  $11$ \\ \hline
$[12,14]$ & $1$ & $1$ &  $1408$ &  $1110$  &  $5$ &  $13$ \\ \hline
$[14,16]$ & $1$ & $0$ &  $1400$ &  $1116$  &  $6$ &  $0$ \\ \hline
$[16,18]$ & $4$ & $0$ &  $1391$ &  $1068$  &  $13$ &  $1$ \\ \hline
$[18,20]$ & $6$ & $0$ &  $1395$ &  $1099$  &  $18$ &  $1$ \\ \hline
$[20,22]$ & $14$ & $1$ &  $1306$ &  $1024$  &  $33$ &  $2$ \\ \hline
$[22,24]$ & $19$ & $0$ &  $957$ &  $537$  &  $35$ &  $1$ \\ \hline
\hline
\end{tabular} 
\caption{Number of over-expressed (second column) and under-expressed (third column) navigation point pairs observed in the 334 AIRAC during different time windows of the day. The fourth and fifth column show the number of navigation point pairs with at least one or five planned flights, respectively. The fifth and sixth columns indicate the number of OE and UE observed in at least one of the 5 AIRACs from 334 to 338. Data shown in this table were obtained by using the FDR correction for multiple test comparison with $N_{pair}=146112$. Data refer to the German (ED) airspace.} \label{fig:EDtempOEpair}
\end{table}

\begin{figure} [H]
\centering 
                 \includegraphics[width=0.90\textwidth]{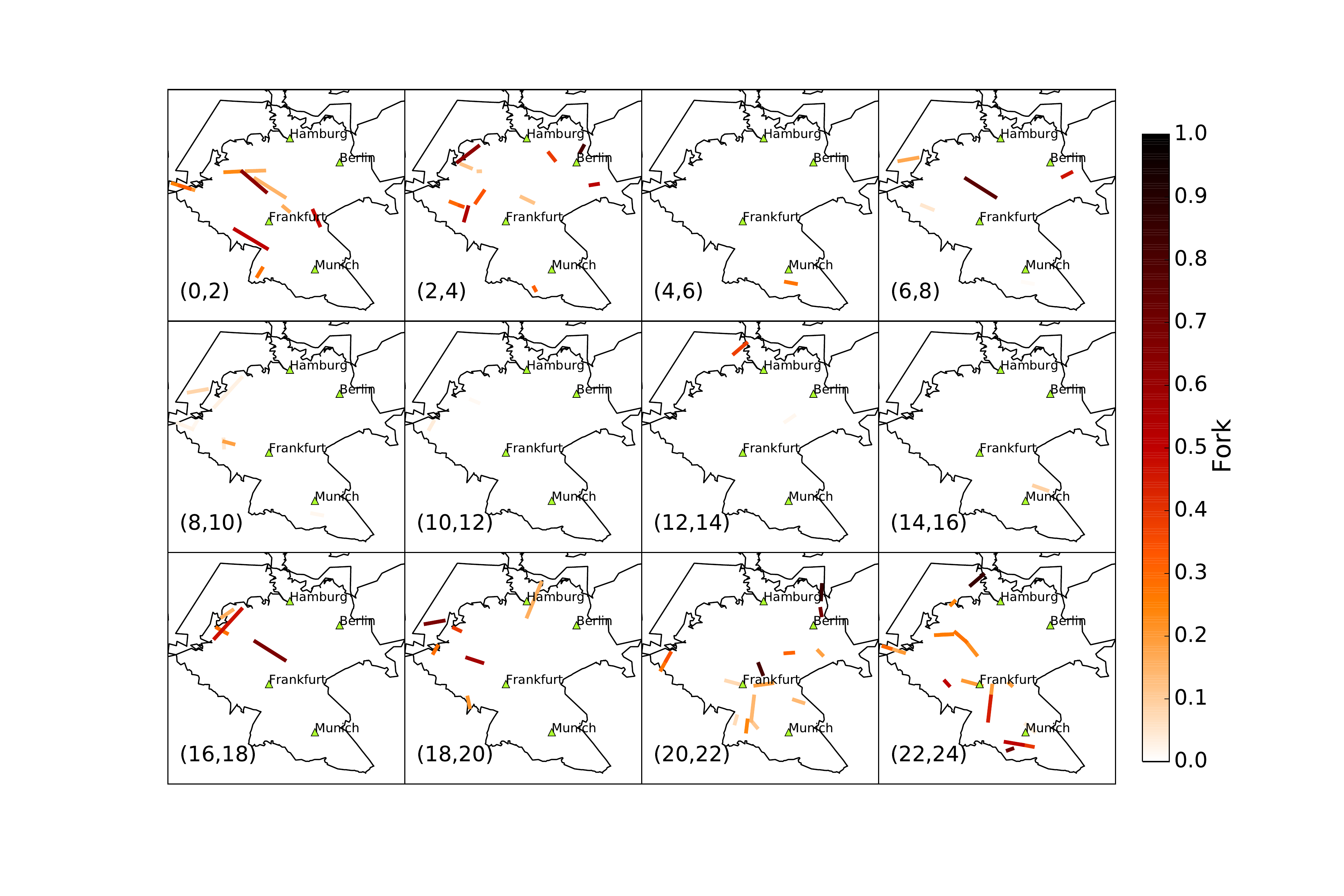}
                 \caption{Spatial localization of the under-expressed pairs and over-expressed pairs over the day. The different  colors are proportional to the measured di-fork value in the considered time-window.} \label{fig:localizationFN3}
\end{figure}

For illustrative purposes, in Fig. \ref{fig:localizationFN3} we show the localization of the under-expressed pairs and over-expressed pairs in the 12 bi-hourly time-windows occurring in a day. The different  colors are proportional to the measured di-fork value in the considered time-window, according to the color code on the right of the figure. Although there seem to be a predominance of segments with colors belonging to upper and lower part of the color bar, in some case we can also see, see panel (2, 4) for instance, some segment with colors belonging to the central part of the color bar. Once again, this indicates that the statistical validation procedure associated to the hypergeometric distribution selects the pairs where the occurrence of deviations is statistically different from the daily average, rather than  pairs with higher or lower di-fork values.

\begin{figure} [H]
\centering 
                 \includegraphics[width=0.50\textwidth]{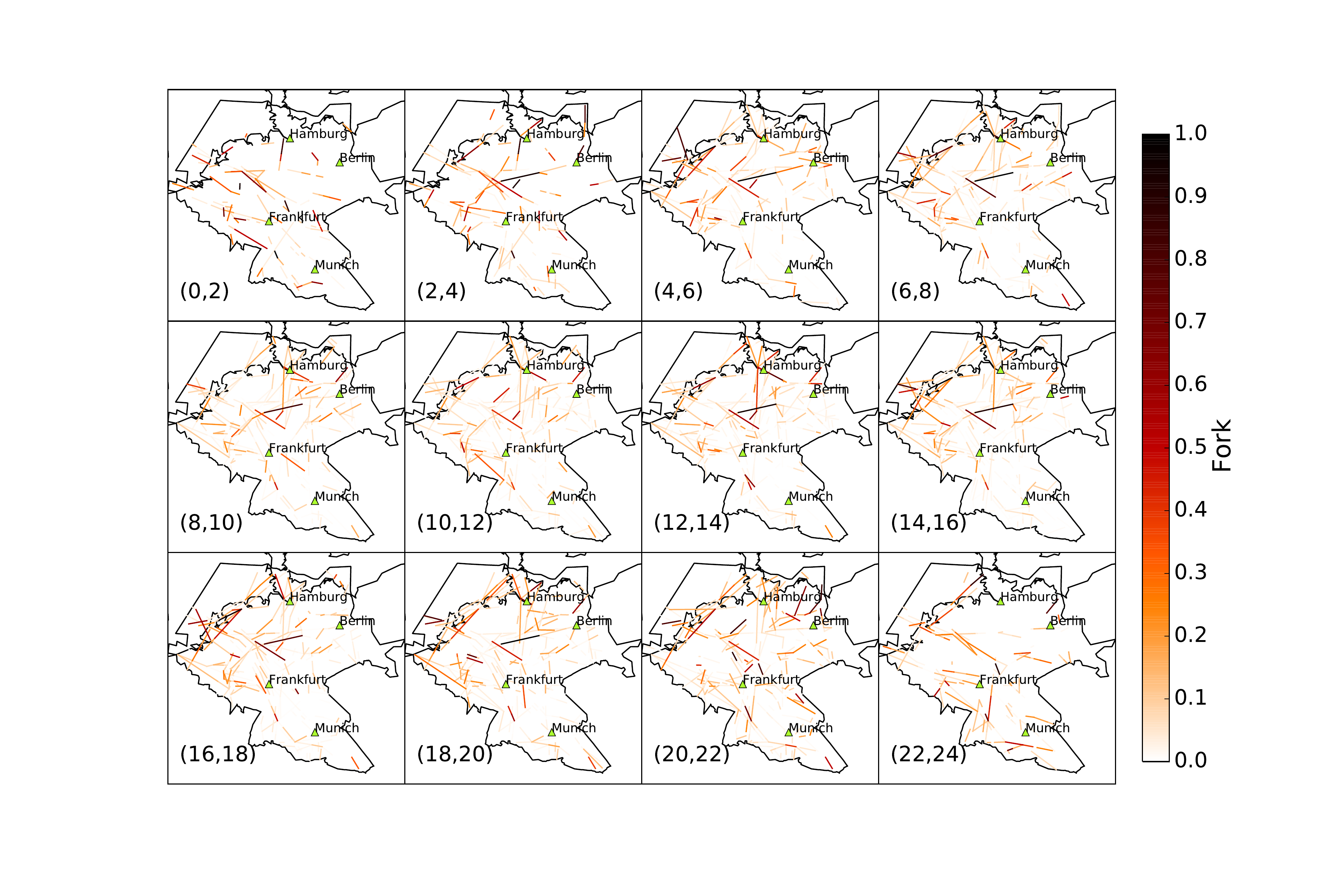}
                 \hspace{-1.truecm}
                 \includegraphics[width=0.50\textwidth]{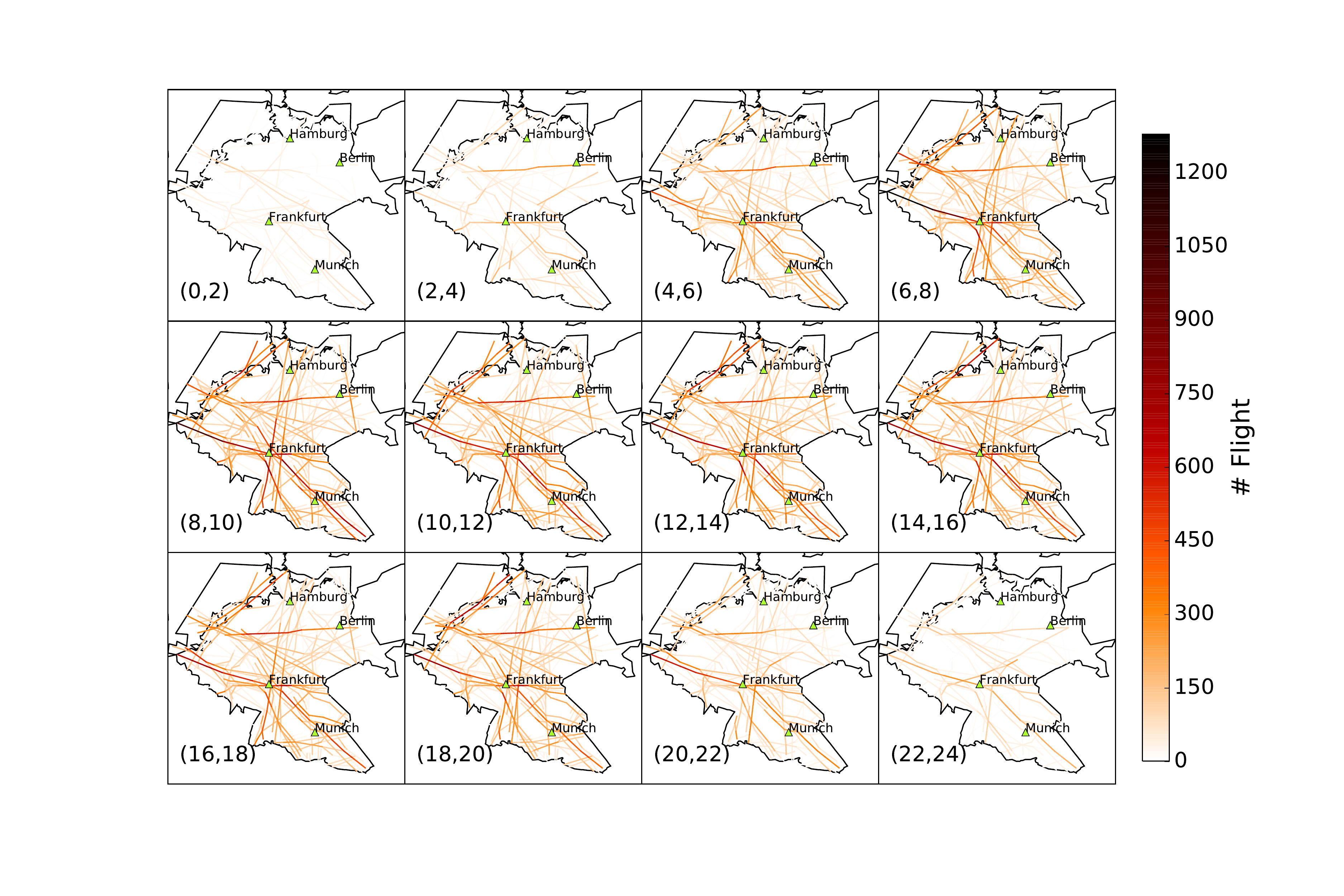}
                 \caption{Spatial localization of the navigation point pairs over the day. In the left panel the different colors are proportional to the measured di-fork value in the considered time-window. In the right panel the different colors are proportional to  the number of aircraft traveling across the navigation point pair in the considered time window.} \label{fig:localizationFN4}
\end{figure}

In fact, in the left panel of Fig. \ref{fig:localizationFN4} we show all navigation point pairs of the considered airspace in the 12 bi-hourly time-windows occurring in a day. As in the above case, the different colors are proportional to the measured di-fork value in the considered time-window, according to the color code on the right of the panel\footnote{In this graphical representation the navigation point pairs $C=(a,b)$ and $\overline{C}=(b,a)$ are considered together. In other words, the di-fork value represented here is defined as  $( DF(C) + DF(\overline{C})) / ( PF(C) + PF(\overline{C}) )$. Moreover, we only show the segments such that $( PF(C) + PF(\overline{C}) ) \ge 5$}. The comparison of such panel with Fig. \ref{fig:localizationFN3} clearly shows that navigation point pairs with larger di-fork values are not over-expressed as well as navigation point pairs with small di-fork values are not under-expressed. In the right panel of Fig. \ref{fig:localizationFN4} we show again all navigation point pairs of the considered airspace in the 12 bi-hourly time-windows occurring in a day. However, here the colors are proportional to  the number of aircraft traveling across the navigation point pair in the considered time window\footnote{In this graphical representation the navigation point pairs $C=(a,b)$ and $\overline{C}=(b,a)$ are considered together. Accordingly, the number of aircraft shown in the figure is the sum of the number go flights traveling from one navigation point to the other in both directions.}. By comparing such panel with Fig. \ref{fig:localizationFN3} one can clearly see that highly travelled segments are not necessarily over-expressed while poorly  travelled segments are not necessarily under-expressed.

\begin{figure}[H]
\centering
                 \includegraphics[width=0.45\textwidth]{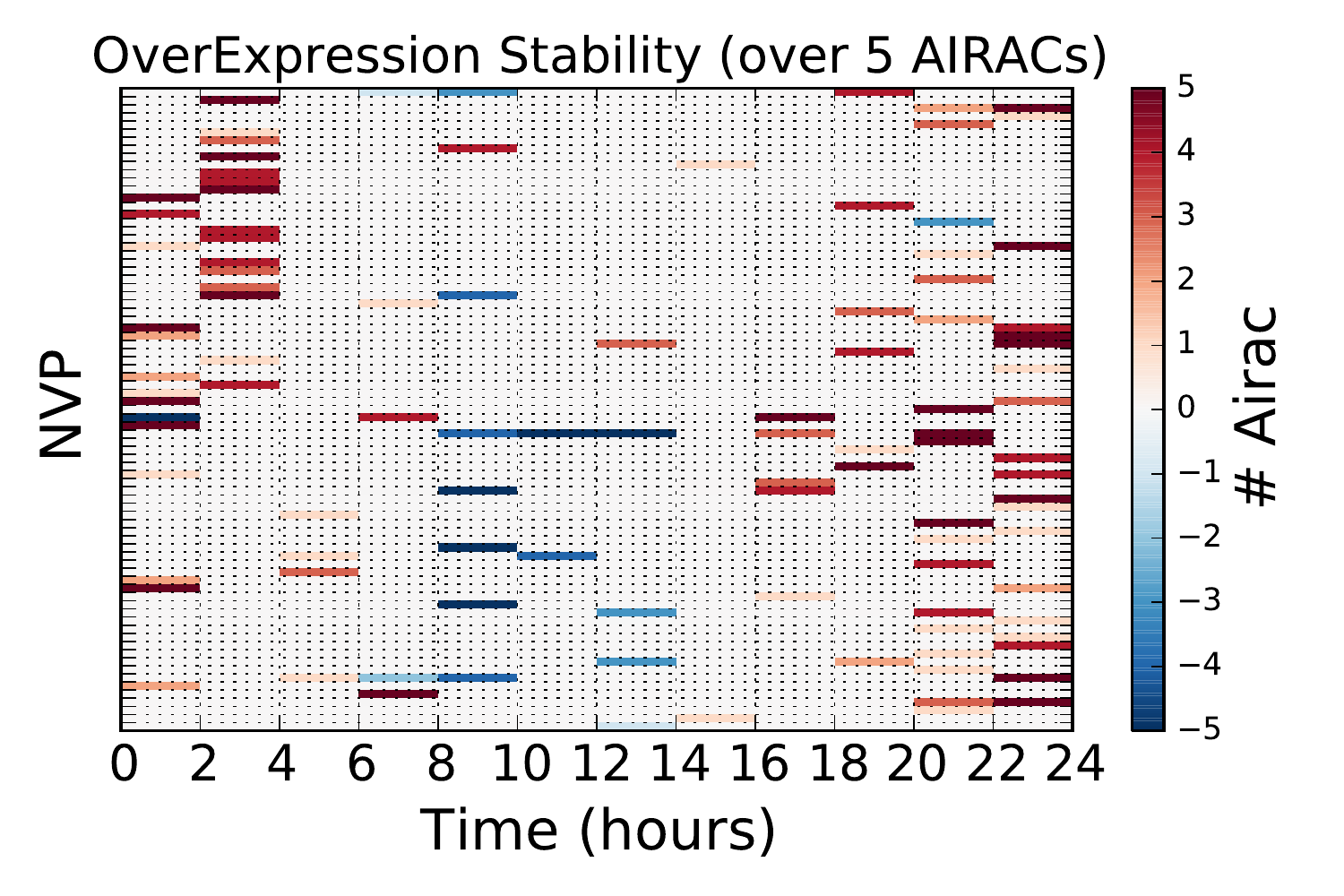}
                 \caption{Color code summary of the persistence of over-expressed and under-expressed navigation point pairs statistically validated during AIRAC 334 and during the successive 4 AIRACs. On the $x$ axis we report the time window of the day and each horizontal line parallel to the $x$ axis represents a navigation point pair statistically validated at least once during AIRAC 334. Positive values (red cells) indicate that the navigation point pair is repeatedly over-expressed during different AIRACs at the considered time window. Negative values (blue cells) indicate repeated under-expression. Data refer to the five AIRACs from 334 to 338 and to navigation point pairs of the German (ED) airspace. Data shown in this table were obtained by using the FDR correction for multiple test comparison with $N_{pair}=146112$.} \label{fig:stability}
\end{figure}

The spatial localization of the over-(under-)expressed navigation point pairs might change for different time-intervals. Again this is not surprising because the di-fork metric not only takes into account the topology of the navigation point network which has a slow dynamics over the day, but it also takes into account the flow of aircraft over the network, which is instead pretty variable over the day. However, stable patterns can be detected. In fact, we investigated the temporal persistence of the over-expression and under-expression of flight deviations at navigation point pairs in Fig. \ref{fig:stability}. This is a color coded figure showing for each investigated time window of the day and for each statistically validated navigation point pairs of the AIRAC 334 whether each navigation point pair turns out to be also over-expressed or under-expressed in the 4 AIRACs successive to AIRAC 334. In the figure a value of 5 (labeled as a red cell) indicates that the navigation point pair is over-expressed in all five considered AIRACs. Negative numbers indicate under-expression. It is worth mentioning that a pair over-expressed (under-expressed) in a certain  AIRAC never happens to be under-expressed (over-expressed) in the other 4 AIRACs. For comparison, in the fifth and sixth column of Table \ref{fig:EDtempOEpair} we indicate the number of OE and UE observed in at least one of the 5 AIRACs from 334 to 338. One can see that there are only a few navigation point pairs that are over-expressed during the 5 AIRACs both during day time and during night-time. The two periods of the day show a quite different general pattern suggesting again different underlying reasons for the deviations of the planned flight trajectories. 
In fact, the figure is consistent with our previous findings and gives us more information. It seems that during the night, controllers are consistently deviating some of the flows in the airspaces, probably to shorten the corresponding trajectories. During the day on the other hand, controllers are stabilizing the horizontal deviations, especially some flows, which are probably more complex than the others. This also complements the results of \cite{ATMgerald}, in which we were only able to see that some navigation points were consistently over-used for deviations. 

As a result, the di-fork metric seems able to show in quantitative way that the fraction of deviations occurring at the level of single navigation point pairs during the day is not a random variable. Rather it follows patterns that reveal to be stable over different AIRACs. Therefore it provides an useful instrument for a ``microscopic'' statistical characterization of the deviations from planned flight trajectories in the air traffic management procedures.

\section{Conclusions} \label{concl}

The air transportation system is a complex system with a pronounced intraday dynamics and a marked spatial heterogeneity. For example, by using traffic data, we have verified that day flights and night flights present a different value for their average length and a different distribution of flight length. In the air traffic management procedures, the interaction between pilots and air traffic controllers is primarily devoted to conflict resolutions aiming to prevent safety problems. Once safety problems are positively avoided or solved the interaction between pilots and air controllers focuses on possible improvement of the efficiency of the air transportation system. To analyze the effect on the efficiency improvement of the system of their interaction we have considered a very simple measure of flight efficiency based on the comparison between the length of planned or realized flight trajectory with the grand circle distance. 
Our results show  that night-time flights (in particular during the time interval from 8:00 pm to 4:00 am) are on average more efficient than day-time flights. Moreover, the gain of average efficiency obtained in the realized trajectories is systematically larger during night-time. Our results show an asymmetry in the change of efficiency of flights during night-time. Specifically for large changes in absolute value the improvement is more evident than the deterioration. This asymmetry is not detected during day-time when efficiency improvement is quite balanced by efficiency deterioration.

The strategic interaction present between pilots and air controllers is also reflected in the observation that flight trajectory deviations preferentially occur near the origin rather than close to destination of the flight. Moreover, flight trajectory deviations occur at an angle-to-destination that is a non monotonic function of the angle with a maximum observed close to 20 degree. Pilots and air controllers are most probably solving on average different kind of problems during day and night. In fact we observe that the fraction of flight trajectory deviations is higher during night-time than during day-time intraday time windows. We also detect that the fraction of flight trajectory deviations is an inverse function of the number of flights observed in the investigated time window.

Our study shows that the time of the day plays an important role in setting the most probable type of interaction between pilots and air controllers. Indeed, in addition to time there is also a role of the specific geographical location of the considered navigation point pair. To clarify this point we introduce a new metric called di-fork that is useful to track the trajectory deviations at the level of single navigation point pairs. By making use of this metric, we can detect the set of navigation point pairs presenting a number of flight trajectory deviations that are over-expressed or under-expressed with respect to a statistical null hypothesis assuming $(i)$ that deviations occur randomly over the day and $(ii)$ taking into account the heterogeneous number of flights planned to fly through the navigation point pair over the day. The detected set of over-expressed and under-expressed navigation point pairs is persistent over a time period spanning 5 successive AIRACs, i.e. for up to at least 140 days. This result quantitatively shows that the fraction of deviations occurring during the day is not a random variable. Rather, it corresponds to the effort of making the system more efficient under certain constraints due, for example, to safety and capacity issues. 

We believe our results present a clear statistical evidence of the ability of the air traffic management system of on average improving air traffic performances with respect to the set of planned flight trajectories. The improvement of performances are relatively more evident during night-time time windows when constraints related to the capacity of air sectors are less stringent. 

In summary our results show that the interaction between pilots and air controllers is a complex one that that is the result of a learning process aiming not only at the prevention and resolution of safety problems but also to the improvement of the performances of each single airline and of the entire air transportation system.
  
\section*{Acknowledgements}

This work was co-financed by EUROCONTROL on behalf of the SESAR Joint Undertaking in the context of SESAR Work Package E project ELSA ``Empirically grounded agent based model for the future ATM scenario''. FL acknowledges support by the European Community's H2020 Program under the scheme ÔINFRAIA-1-2014-2015: Research InfrastructuresÕ, grant agreement \#654024 ÔSoBigData: Social Mining \& Big Data EcosystemÕ (http://www.sobigdata.eu).



\begin{thebibliography}{99}

\bibitem{ZaninLillo13}
Zanin M, Lillo F,
``Modelling the air transport with complex networks: A short review'',
\newblock Eur Phys J Spec Top 215, 5-21 (2013).

\bibitem{CWbook}
Cook A,  Rivas D (Eds.),
``Complexity science in air traffic management'',
\newblock Ashgate Publishing Limited, England, (2016), ISBN  978-1-4724-6037-0.


\bibitem{AIRP1}
Wang J, Mo H, Wang F, Jin F, 
``Exploring the network structure and nodal  centrality of China's air transport network: A complex network approach'',
\newblock J Transp Geogr 19, 712-721 (2001).

\bibitem{AIRP2}
Li-Ping C, Ru W, Hang S, Xin-Ping X, Jin-Song Z, et~al.,
``Structural  properties of US flight network'',
\newblock Chin Phys Lett 20, 1393-1396 (2003).

\bibitem{AIRP3}
Barrat A, Barth\'{e}lemy M, Pastor-Satorras R, Vespignani A, 
``The  architecture of complex weighted networks'',
\newblock PNAS 101, 3747-3752 (2004).

\bibitem{AIRP4}
Chi L-P, Cai X,
``Structural changes caused by error and attack tolerance in US airport network'',
\newblock International Journal of Modern Physics B 18,  2394-2400 (2004).

\bibitem{AIRP5}
Li W, Cai X, 
``Statistical analysis of airport network of China'',
\newblock Phys Rev E 69, 046106 (2004).

\bibitem{AIRP6}
Guimer\`{a} R, Mossa S, Turtschi A, Amaral L, 
``The worldwide air  transportation network: Anomalous centrality, community structure, and  cities' global roles'',
\newblock PNAS 102, 7794-7799 (2005).

\bibitem{AIRP7}
Colizza V, Barrat A, Barthelemy M, Vespignani A, 
``The role of the airline transportation network in the prediction and predictability of global epidemics'', 
\newblock PNAS 103, 2015-2020 (2006).

\bibitem{AIRP8}
Guida M, Maria F, 
``Topology of the Italian airport network: A scale-free  small-world network with a fractal structure?'',
\newblock Chaos Solitons Fractals 31, 527-536, (2007).

\bibitem{AIRP9}
Bagler G, 
``Analysis of the airport network of India as a complex weighted  network'',
\newblock Physica A 387, 2972-2980 (2008).

\bibitem{AIRP10}
Xu Z, Harriss R, 
``Exploring the structure of the U.S. intercity passenger  air transportation network: a weighted complex network approach'',
\newblock GeoJournal 73, 87-102 (2008).

\bibitem{AIRP11}
Cardillo A, G\'omez-Gardenes J, Zanin M, Romance M, Papo D, del Pozo F, Boccaletti S,
``Emergence of network features from multiplexity'',
\newblock Sci. Rep. 3, 1344 (2013).

\bibitem{AIRP12}
Gomes MFC, Pastore y Piontti A, Rossi L, Chao D, Longini I, Halloran ME, Vespignani A,
``Assessing the International Spreading Risk Associated with the 2014 West African Ebola Outbreak'',
\newblock PLOS Currents Outbreaks, 2014 Sep 2, (2014).



\bibitem{ATM1}
Malighetti P, Paleari S, Redondi R,
``Connectivity of the European airport network: ÔÔSelf-help hubbingÕÕ and business implications'',
\newblock Journal of Air Transport Management 14, 53-65, (2008).

\bibitem{ATM2}
Lacasa L, Cea M, Zanin M,
``Jamming transition in air transportation networks'',
\newblock Physica A 388, 3948-3954 (2009).

\bibitem{ATM3}
Ben Amor S, Bui M,
``A Complex Systems Approach in Modeling Airspace Congestion Dynamics'',
\newblock Journal of Air Transport Studies, 3 (1), 39-56 (2012).

\bibitem{ATM4}
Cai K, Zhang J, Du W, Cao X, 
``Analysis of the Chinese air route network as  a complex network'',
\newblock Chin Phys B 21, 028903 (2012).

\bibitem{ATM5}
Cook A, Tanner G, Crist\'obal S, Zanin M,  
``New perspectives for air transport performance'',
\newblock in: Proceedings of the Third SESAR Innovation Days, Schaefer D (Ed.), Stockholm, (2013).

\bibitem{ATM6}
Fleurquin P, Ramasco J J, Eguiluz V M,
``Systemic delay propagation in the US airport network'',
\newblock Scientific Reports 3, 1159 (2013).

\bibitem{ATM7}
Pyrgiotis N, Malone KM, Odoni A,
``Modelling delay propagation within an airport network'',
\newblock Transportation Research Part C 27, 60-75 (2013).

\bibitem{ATM8}
Cardillo A, Zanin M, G\'omez-Gardenes J, Romance M, del Amo AJG, Boccaletti S,  
``Modeling the multi-layer nature of the European air transport network: resilience and passengers re-scheduling under random failures'', 
\newblock Eur Phys J Spec Top 215, 23-33 (2013).

\bibitem{ATM9}
Campanelli B, Fleurquin P, Eguiluz VM, Ramasco JJ, Arranz A, Etxebarria I, Ciruelos C, 
``Modelling reactionary delays in the European air transport network'',
\newblock in: Proceedings of the Fourth SESAR Innovation Days, Schaefer D (Ed.), Madrid, (2014).

\bibitem{ATM10}
Zanin M,
``Network analysis reveals patterns behind air safety events'',
\newblock Physica A 401, 201-206 (2014).

\bibitem{ATM11} 
Blom HAP, Cook A, Lillo F, Mantegna RN, Miccich\`e S, Rivas D, V\'azquez R, Zanin M, 
``Applying complexity science to air traffic management'', 
\newblock JATM 42, 149-158 (2015).


\bibitem{plos} 
Gurtner G, Vitali S, Cipolla M, Lillo F, Mantegna RN, Miccich\`e S, Pozzi S, 
``Multi-Scale Analysis of the European Airspace Using Network Community Detection'', 
\newblock PLoS ONE {\bf 9}  e94414 (2014).

\bibitem{conops} 
SESAR, ``SESAR Concept of Operations Step 1'', 2012.

\bibitem{DDR} 
EUROCONTROL,
"DDR Reference Manual 1.5.8, DDR Version: 1.5.8. (2010)
{\tt{http://www.eurocontrol.int/services/demand-data-repository-ddr}}

\bibitem{NEVAC}
{\tt{http://www.eurocontrol.int/eec/public/standard\_page/NCD\_nevac\_home.html}}

\bibitem{PRR}
Performance Review Commission, Performance Review Report 2013.
\newblock {\tt{http://www.eurocontrol.int/sites/default/files/}}
{\tt{publication/files/prr-2013.pdf}}

\bibitem{Wilson} 
Wilson EB,
``Probable inference, the law of succession, and statistical inference'', 
\newblock Journal of the American Statistical Association 22, 209-212 (1927).
 
\bibitem{ATMgerald}
Bongiorno C, Miccich\`e S, Mantegna RN,  Gurtner G, Lillo F, Pozzi S, 
``Adaptative air traffic network: statistical regularities in air traffic management'', 
\newblock presented at the 11$^{th}$ USA/Europe ATM R\&D Seminar, 23-26 June 2015, Lisbon, Portugal. 
{\tt{http://www.atmseminar.org/seminarContent/seminar11/papers/440\_Gurtner\_0126150101-Final-Paper-4-30-15.pdf}}

\bibitem{FI1} 
Fisher R A, 
On the interpretation of $\chi^2$ from contingency tables, and the calculation of p. 
\newblock Journal of the Royal Statistical Society , 87Ð94 (1922).


\bibitem{FIHYP} 
Rivals I, Personnaz L, Taing L, and Potier M-C, 
Enrichment or depletion of a go category within a class of genes: which test? 
\newblock Bioinformatics 23 (4), 401Ð407 (2007).

\bibitem{benj} 
Benjamini Y, Hochberg Y,
``Controlling the false discovery rate: a practical and powerful approach to multiple testing'',
\newblock J R Statist Soc B 57, 289-300 (1995).

\bibitem{garla} 
Squartini T, Picciolo F, Ruzzenenti F, Garlaschelli D,
``Reciprocity of weighted networks'',
\newblock Sci. Rep. 3, 2729 (2013).
 
\end{thebibliography}
\end{document}